\documentclass[10pt]{article}
\usepackage[utf8]{inputenc}
\usepackage[english]{babel}
\usepackage[round]{natbib}
\usepackage{amsmath}
\usepackage{times}
\usepackage[title]{appendix}
\usepackage{amssymb}
\usepackage{ulem}
\usepackage{color}
\usepackage{multicol,caption}
\usepackage{graphicx}
\usepackage{titlesec}
\usepackage{subfigure}
\usepackage{verbatim}
\usepackage{float}
\usepackage{titlesec}
\usepackage{hyperref}
\usepackage{bm}
\makeatletter
\def\UrlAlphabet{%
      \do\a\do\b\do\c\do\d\do\e\do\f\do\g\do\h\do\i\do\j%
      \do\k\do\l\do\m\do\n\do\o\do\p\do\q\do\r\do\s\do\t%
      \do\u\do\v\do\w\do\x\do\y\do\z\do\A\do\B\do\C\do\D%
      \do\E\do\F\do\G\do\H\do\I\do\J\do\K\do\L\do\M\do\N%
      \do\O\do\P\do\Q\do\R\do\S\do\T\do\U\do\V\do\W\do\X%
      \do\Y\do\Z}
\def\UrlDigits{\do\1\do\2\do\3\do\4\do\5\do\6\do\7\do\8\do\9\do\0}
\g@addto@macro{\UrlBreaks}{\UrlOrds}
\g@addto@macro{\UrlBreaks}{\UrlAlphabet}
\g@addto@macro{\UrlBreaks}{\UrlDigits}
\makeatother
\titleformat*{\section}{\large\bfseries}
\titleformat*{\subsection}{\normalsize\bfseries}
\titleformat*{\subsubsection}{\normalsize\bfseries}
\usepackage[paperwidth=8.5in, paperheight=11in,margin=1 in]{geometry}
\title{\fontsize{18}{36}\selectfont \textbf{Full Resolution of Extreme Ship Response Statistics}}
\author{\fontsize{14}{36}\selectfont Xianliang Gong, Zhou Zhang, Kevin Maki, Yulin Pan \\
\fontsize{14}{36}\selectfont (Department of Naval Architecture and Marine Engineering, \\
\fontsize{14}{36}\selectfont University of Michigan, USA)}

\date{}
\usepackage{fancyhdr}

\begin{document}
\maketitle
\thispagestyle{fancy}
\begin{multicols}{2}

\section*{\selectfont ABSTRACT}
We consider the statistics of extreme ship motions in a nonlinear irregular wave field. While an accurate computation is possible by using a full Monte-Carlo method to cover all individual wave conditions, the computational cost may become prohibitively high (when coupled with high-fidelity simulations) due to the rareness of the extreme events. In this work, following existing methods of sequential sampling and wave group parameterization, we implement a framework incorporating nonlinear wave simulation and ship response CFD simulation, which allows the extreme ship motion statistics in nonlinear wave field to be computed efficiently. We test the validity of the framework for the cases of ship response calculated by a nonlinear roll equation, and show the importance of wave nonlinearity to the extreme response statistics. Finally, the framework is coupled with the CFD model to demonstrate its applicability to more realistic and general ship motion problems. 

%applied to study a ship roll problem in which the response is simulated by CFD.

%This development leverages a range of physics and learning based approaches, including nonlinear wave simulations, ship response simulations (e.g., CFD), dimension-reduction techniques, Gaussian process regression (Kriging). 

\setlength{\parindent}{0.5in}

\section*{INTRODUCTION}
 Ship motions are excited in a stochastic environment of ocean waves. The extreme ship motions, usually significantly larger than the statistical average, can cause severe damage to ships. Reliable quantification and physical understanding of these extreme motion statistics is of vital importance to the survivability of ships,  especially in high-sea states. 
 
 The extreme ship motions can be caused by different physical wave conditions. While extreme waves are generally recognized as an important factor, extreme motions can also be induced in moderate wave conditions through nonlinear wave-body interaction mechanisms. These wave-body interactions may lead to parametric roll resonance, surf-riding, broaching and other dynamical phenomena. A complete dynamical model for the extreme ship motion needs to account for the nonlinearity in both wave field and wave-ship interactions. With the increase of the computational power, it is now possible to simulate an individual extreme event with high fidelity (e.g., through the potential flow simulation for nonlinear wave fields and CFD for wave-ship interactions). However, the resolution of the extreme motion statistics requires a Monte-Carlo method of running CFD to sample the ensemble of all nonlinear wave realizations. This can become computationally intractable for irregular waves given the rarity of extreme events and the high-dimensional space of the wave field, which require a prohibitively large number of CFD simulations.

 %such as wave-body resonance and parametric excitation. 

% wave nonlinear 
% response nonlinear 
% need CFD here 

To obtain a practical solution of extreme ship motion statistics, earlier method \citep{soding1986computing} relies on linear extrapolation between $ln(T)$ and $1/H_s^2$, with $T$ the failure return period and $H_s$ the significant wave height. However, this is later shown to result in under-estimation of $T$, i.e., conservative results \citep{shigunov2016probabilistic}. More recent (and robust) approaches mainly rely on the principle of separation \cite[see review in][]{belenky2012approaches}, which splits the computation into rare (R) and non-rare (NR) sub-problems.  

Within the category of R-NR approaches, the (Envelope) Peaks-Over-Threshold (POT or EPOT) method \citep{mctaggart2000ship,campbell2016application,weem2019senvelope} identifies the upcrossing rate over a moderate threshold $S_m$ in the NR problem, and extrapolates the distribution to a larger threshold based on the asymptotic extreme value theory in the R problem; the split-time method \citep{belenky2016split,belenky2018tail} identifies the upcrossing rate and distribution of derivative process (say roll rate) at $S_m$ in the NR problem, and computes the probability of failure conditioning on the derivative process at $S_m$ in the R problem; The critical wave group approach \citep{themelis2007probabilistic,anastopoulos2016towards,anastopoulos2016ship,anastopoulos2019evaluation} uses the NR problem to evaluate the distribution of ship initial conditions encountering a wave group, and relates the extreme statistics to the probability of critical wave groups based on the Markov process of wave crest in the R problem. All three methods achieve certain levels of success. In particular, \cite{belenky2018tail} addresses the validity of EPOT and split-time methods; Positive results are reported for the critical wave group approach \citep{malara2014maximum,shigunov2019critical}, with its experimental implementation discussed in \cite{bassler2010application,bassler2010characteristics,bassler2019experimental,anastopoulos2016towards}.
%Another set of methods, based on the design load generator \citep{alford2009generating,kim2014statistical,xu2020method}, can be used to generate short-time wave conditions leading to the extreme ship motions. 

 %The available approaches on the computation of the extreme event statistics include extreme value extrapolation \citep{mctaggart2000ship}, envelope peaks over threshold method \citep{belenky2011evaluation}, design loads generator \citep{kim2014statistical}, and large deviation theory \citep{touchette2009large,dematteis2018rogue}. The extreme value extrapolation has been widely used and can be applied without a prior knowledge on the physical process. However, the analysis assumes the independent and identically distributed response, which is not the truth for many situations. The design loads generator method can efficiently construct an ensemble of short time histories reproducing the extreme statistics, but requires the assumption of Gaussian process on the response process. The large deviation theory needs a costly optimization process to find the feature that dominate the extreme PDF, but the resulting distributions cannot always capture the nontrivial shape of the tail \citep{mohamad2018sequential}.

\begin{figure*}
\begin{center}
  \includegraphics[width=14cm]{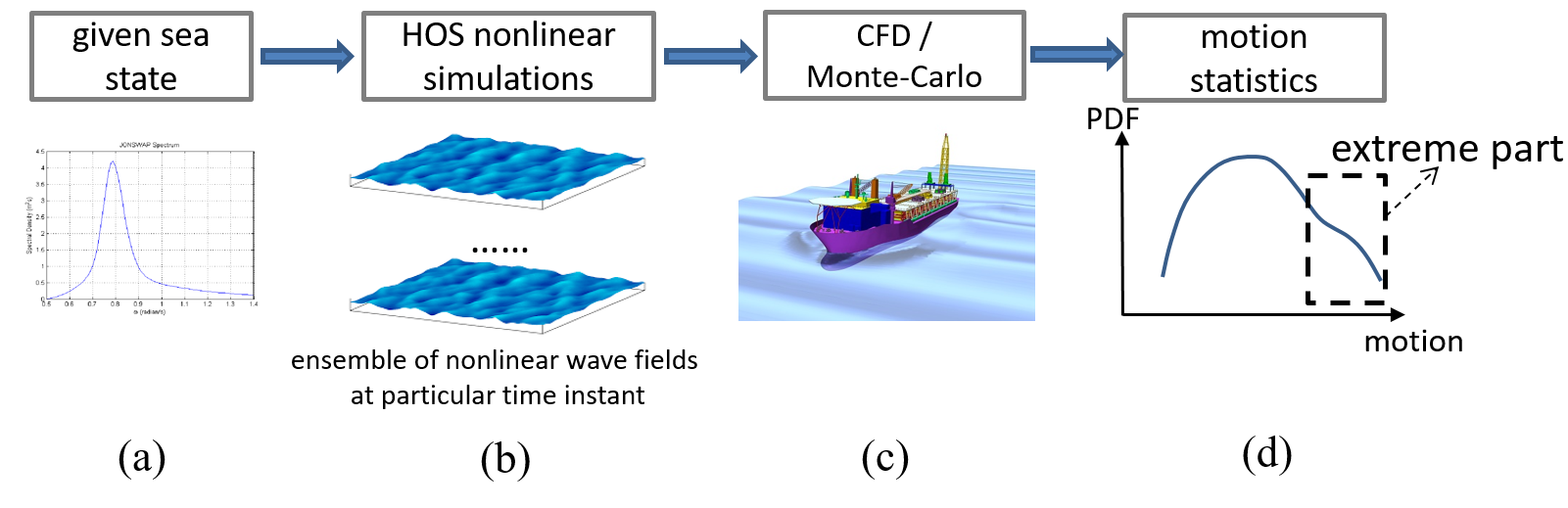}
 \end{center}
 \caption{ \label{fig:fn:1}
  Outline of the full (brute-force) framework to compute the extreme PDF of ship motion}
 \end{figure*}
% including CFD 
 
In spite of the success of previous methods, an efficient method which incorporates the nonlinearity from both wave field and wave-body interactions is still lacking. In this paper, following recent methods of wave group parameterization \citep{cousins2016reduced} and sequential sampling \citep{mohamad2018sequential, hu2016global, echard2011ak}, we implement a framework which enables an efficient resolution of the statistics of extreme ship responses in irregular nonlinear wave fields. In addition to some algorithmic improvement on the existing methods, our framework allows the wave nonlinearity to be captured through ensemble simulations by high-order spectral method \citep{dommermuth1987high, west1987new}. We restrict the application of the framework to two-dimensional narrow-band wave field and define the ship motion statistics in terms of the maximum motion response in each (isolated) wave group (hereafter group-maximum motion statistics). The developed framework is benchmarked for its effectiveness in accurately resolving the group-maximum statistics in a problem where a roll equation is used to compute the group-maximum motion response. The effects of wave nonlinearity on extreme response statistics are illustrated in an example with an evolving nonlinear wave field. We finally demonstrate the coupling of the framework with CFD simulations to study the realistic roll-motion statistics of a two-dimensional square-shaped hull.

% In this work, 

\section*{COMPUTATIONAL FRAMEWORK}
Given a narrow-band (but otherwise arbitrary) wave spectrum and a certain ship geometry, the purpose of our computation is to resolve the probability distribution function (PDF) of the group-maximum ship motion with high precision on its tail part. An outline of the full (brute-force) computation is illustrated in Figure \ref{fig:fn:1}. In this process, the procedure of computing the ensemble of nonlinear wave fields from a given spectrum ((a)-(b)) can be accomplished by an ensemble run of the high-order spectral (HOS) method \citep{dommermuth1987high,west1987new} starting from different initial random phase distributions. This is in essence a full Monte-Carlo method, which is computationally viable due to the highly efficient spectral algorithm in HOS. The difficulty in the computation of Figure \ref{fig:fn:1} lies in the process from (b) to (d), where CFD simulations are used to sample the ensemble of nonlinear wave fields to resolve the motion PDF. Due to the high dimensionality of the wave fields and the rarity of the extreme motion response, a large number of CFD simulations are required to obtain converged statistics for the tail of the motion PDF. This can become computationally prohibitive for complex problems where each CFD simulation is already expensive.     

To enable the computation in Figure \ref{fig:fn:1}, efficient algorithms have to be developed to realize the process from (b) to (d). We next describe two key methods for this purpose. In particular, we use a wave group parameterization technique to reduce the dimensionality of a narrow-band nonlinear wave field, and a sequential sampling to reduce the number of samples (in the low-dimensional space) to resolve the extreme motion statistics.

% Should we use the spatial representation ?

\subsection*{Wave Group Parameterization}
Wave groups are structures with successive large waves embedded in random wave fields. The concept has been adopted to describe ocean waves in both deterministic and probabilistic ways. The linear deterministic models include the quasi-determinism (QD) theory \citep{boccotti1997general,boccotti2008quasideterminism} and the rare wave group theory \citep{seyffert2016rare}, which construct the deterministic wave profiles conditioning on the occurrence of extreme surface elevation. The nonlinear wave group model includes the celebrated Peregrine soliton \citep{peregrine1983water} as an analytical solution to the nonlinear Schrödinger equation for narrow-band water waves. Recently, a theory of hydrodynamic instanton \citep{dematteis2019experimental} is also proposed to encompass the two approaches (QD and soliton) for a unified explanation of extreme waves. 

On the probabilistic side, the statistics of wave groups (in terms of height, period and run length) is formulated by \cite{longuet1957statistical} through the spectral moments, and by \cite{kimura1980statistical} through the correlation between the successive waves. In \cite{longuet1984statistical}, it is shown that the results given by the two formulations are consistent for small bandwidth of the spectra. However, the validity of these approaches have not been fully tested for (strongly) nonlinear waves. To account for nonlinearity, it is desirable to parameterize (and compute the statistics of) the wave groups directly from data of surface elevations of a nonlinear wave field. 

For this purpose, we first convert the wave elevation field $\eta(x)$ into an envelope process through the Hilbert transform \citep{shum1984estimates} (see Figure 2(a) for a typical case):
\begin{equation}
    \eta(x)=Re\{\rho(x) e^{i k_0 x+i\phi(x)}\},
    \label{etax}
\end{equation}
where $\rho(x)$ is the envelope process, $k_0$ the carry wavenumber, and $\phi(x)$ the phase modulation. Here, we restrict our method only to sufficiently narrow-band wave field where the low-dimensional structure in figure 2 (in particular the long group in $\rho(x)$ and nearly constant $\phi(x)$ in each group) can be clearly identified. Under this situation, each wave group of the envelope can be approximated by a localized Gaussian function:
\begin{equation}
    \rho_c(x)\sim A \exp \frac{-(x-x_c)^2}{2L^2}
    \label{GauWG}
\end{equation}
where $A$ is the group amplitude, $L$ the length scale and $x_c$ the location of the group, $\rho_c(x)$ the segment of the envelope $\rho(x)$ in the vicinity of $x_c$, corresponding to a wave group.

\begin{center}
    \includegraphics[width=7.5cm]{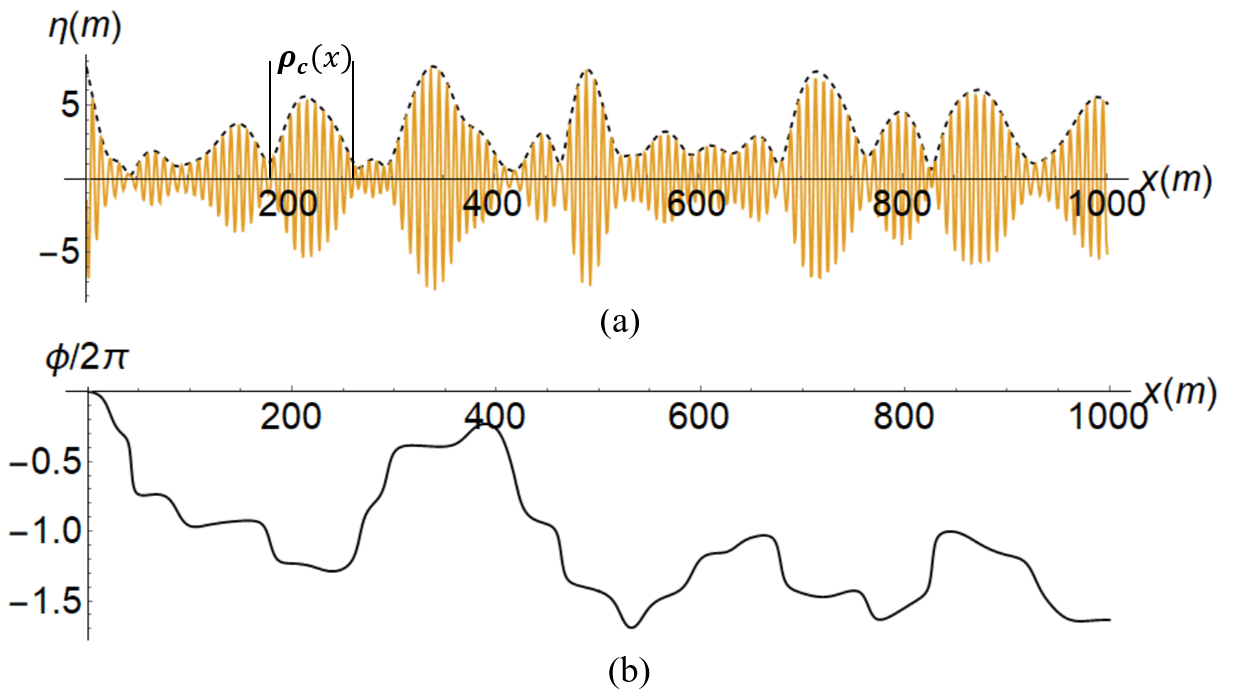}
\captionof{figure}{(a) envelope process $\rho(x)$ (black dashed line) and (b) phase modulation $\phi(x)$ computed from the surface elevation $\eta(x)$ (yellow solid line in (a)) using the Hilbert transform. A localized wave group $\rho_c(x)$ is indicated in (a).}
\label{HT}
\end{center}

%For wave fields with narrow-band spectrum, the wave elevation $\eta(x)$ can be described through the envelope process:
%\begin{equation}
%    \eta(x)=Re\{\rho(x) e^{i k_0 x+i\phi(x)}\},
%    \label{etax}
%\end{equation}
%where $\rho(x)$ is the envelope process, $k_0$ the carry wavenumber, and $\phi(x)$ the phase modulation. Given a realization of surface elevations $\eta(x)$, the wave envelope and phase modulation can be calculated by the Hilbert transform \citep{shum1984estimates} (see Figure 2(a) for a typical case). {\color{red} We remark that while the Hilbert transform can also be applied to broadband wave field, the low-dimensional structure in figure 2 (in particular the group in $\rho(x)$ and nearly constant $\phi(x)$ in each group) can only be clearly (or is much better) identified for a narrow-band wave field}. Under this situation, each wave group of the envelope can be approximated by a localized Gaussian function:
%\begin{equation}
%    \rho_c(x)\sim A \exp \frac{-(x-x_c)^2}{2L^2}
%    \label{GauWG}
%\end{equation}
%where $A$ is the group amplitude, $L$ the length scale and $x_c$ the location of the group, $\rho_c(x)$ the segment of the envelope $\rho(x)$ in the vicinity of $x_c$, corresponding to a wave group.

Our purpose here is to detect $A$ and $L$ for each wave group in the envelope $\rho(x)$, so that a low-dimensional description of the (ensemble) wave field can be established in terms of the joint PDF of $A$ and $L$ ($x_c$ is not important since the ship-motion response does not depend on $x_c$). Our method builds on \cite{cousins2016reduced,cousins2019predicting}, originated from a scale-selection method in the computer vision field for feature detection in images \citep{lindeberg1998feature}. Specifically, the algorithm (see Appendix A for details) finds the local minimum of the normalized second derivative of the space-scale representation function $\bar{S}^{(2)}(x,l)$ (see Eq.\eqref{S2}), and set 
\begin{align}
   (L_0,x_c) & ={\rm{argmin}}_{l,x}\bar{S}^{(2)}(x,l),\nonumber \\
   L &= \frac{1}{\sqrt{2}}L_0.
   \label{opt1}
\end{align}
The second equation of Eq.\eqref{opt1} is obtained from theoretical consideration on the local minimum of $\bar{S}^{(2)}(x,l)$ when $\rho_c(x)$ is in an exact Gaussian shape \citep{lindeberg1998feature,cousins2016reduced}. 
Given $x_c$, $A$ for the same wave group is determined correspondingly as $A=\rho(x_c)$. 

%To explain the reason, we show in Appendix A that the computation of $\bar{S}^{(2)}(x,L)$ is equivalent to a Mexican-hat wavelet transform. However, the wave group is then approximated by a Gaussian function \eqref{GauWG}, which has a scale mismatch with the Mexican-hat wavelet.

While the detection algorithm is simple to implement, we find that the scale $L$ is in general under-predicted. Figure \ref{detect}(a) shows a typical result where this point is elucidated. This is due to the finite length of the actual group $\rho_c(x)$, in contrast to an ideal Gaussian group with infinite length. As a result, the adjacent signal $\rho(x)$ around $\rho_c(x)$ affects the scale resolution. To address this issue, we consider the value of $L$ calculated by \eqref{opt1} as an initial guess, and conduct another optimization problem to directly maximize a similarity measure $C$ (eq.\eqref{Cmeasure} evaluated over 1.3 standard deviation around the peak) between $\rho_c(x)$ and $A \exp [-(x-x_c)^2/(2L^2)]$, and set    
\begin{equation}
    L={\rm{argmax}}_{L} C(L,A,x_{c}).
    \label{opt2}
\end{equation}
We use Newton's iteration method to solve \eqref{opt2}. The result after applying \eqref{opt2} is shown in Figure \ref{detect}(b), where clear improvement (in terms of the closeness between the detected groups and original signal) can be visualized compared to \ref{detect}(a). For this wave field, the average value of $C$ increases from 0.78 to 0.91 after applying \eqref{opt2}.

\begin{center}
    \includegraphics[width=7.5cm]{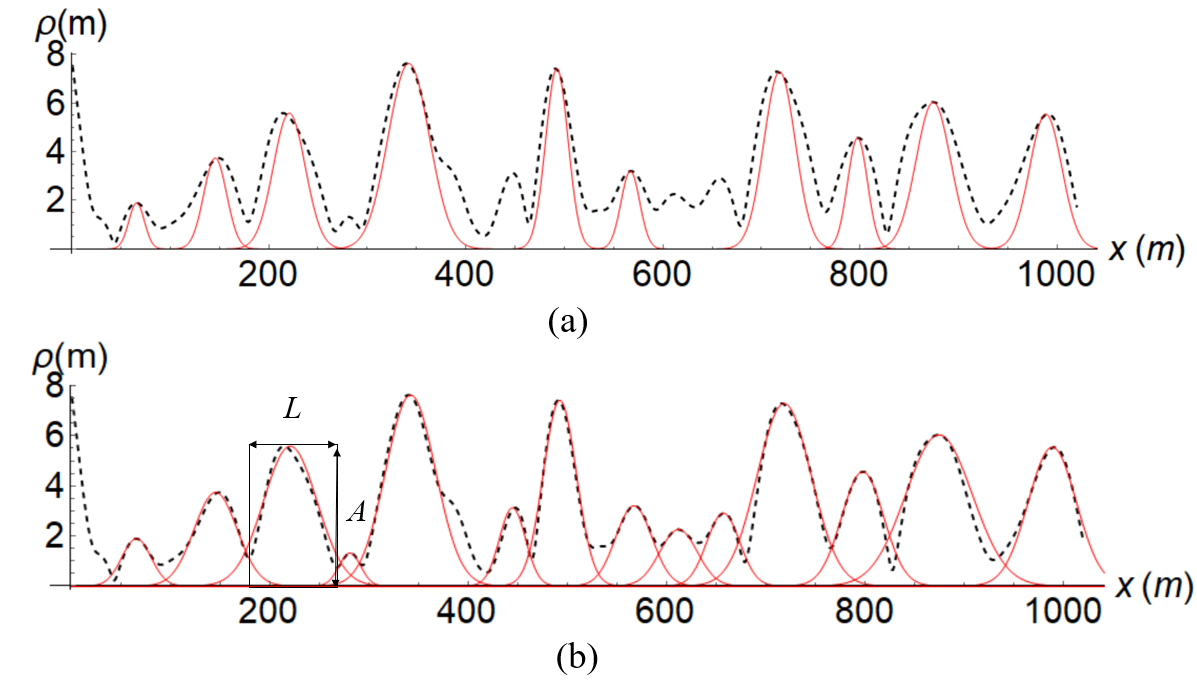}
\captionof{figure}{The localized Gaussian wave groups (red line) calculated using (a) Eqs.\eqref{opt1} and \eqref{crit} and (b) Eqs.\eqref{opt1}, \eqref{opt2} and \eqref{crit}, for an envelope process (black dashed line).}
  \label{detect}
\end{center}

\begin{center}
    \includegraphics[width=6cm]{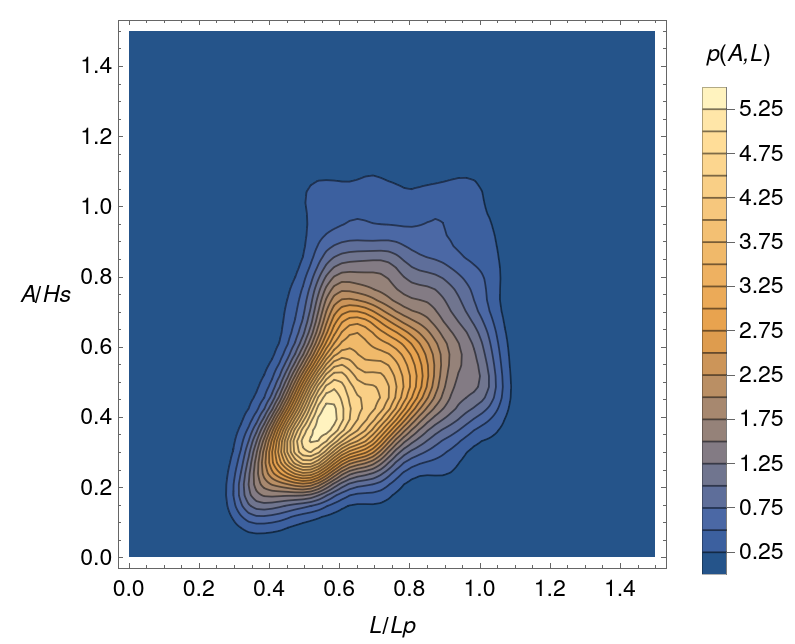}
\captionof{figure}{An example of the joint PDF $p(A,L)$ of $A$ and $L$, normalized respectively by the significant wave height $H_s$ and spectral peak wavelength $L_p$. }
\label{LAPDF}
\end{center}

The group detection algorithm can be applied to the ensemble of nonlinear wave fields to resolve a collection of pairs $L$ and $A$, which are then used to calculate $p(A,L)$, the joint PDF of $L$ and $A$ (see Figure \ref{LAPDF} for an example). This joint PDF provides a low-dimensional probabilistic description of the nonlinear wave fields, and can be sampled as input to the CFD simulations.

%Applying our detection algorithm to wave fields, we can acquire an ensemble of wave groups. Then we apply kernel density estimation to this ensemble to get the joint PDF of group amplitude and group length scale, serving as the low-dimensional probabilistic space.  

\subsection*{Sequential Sampling}
Given $p(\bm{\theta}\equiv A,L)$, the PDF of a ship motion can be computed from sampling the space of $\bm{\theta}$. Mathematically, we consider a map $T$ which maps a wave group parameter $\bm{\theta}$ to the group-maximum ship motion $r$: 
\begin{equation}
    r=T(\bm{\theta}).
\end{equation} 
Our objective is to resolve the PDF of $r$, $p(r)$, with high precision on its tail part. In practice, the computation of map $T$ requires expensive CFD computations. Hence only a limited number of computations can be conducted. Even though the dimension of sample space is low, the required number of CFD simulations is still too large due to the low probability of the extreme motions. 

%For ships in narrow-band wave field, in terms of extreme response statistic, there are two kinds of statistics, the group maximum PDF, which is the statistics with respect to the largest response in groups, and the temporal PDF, which is generated by uniformly sampling in time. In this work we will focus on the group maximum PDF. 

%After parameterization, the originally dynamical problem becomes a static input-output model, meaning that the maximum response in a wave group does not depend on time. 

\begin{center}
    \includegraphics[width=7.5cm]{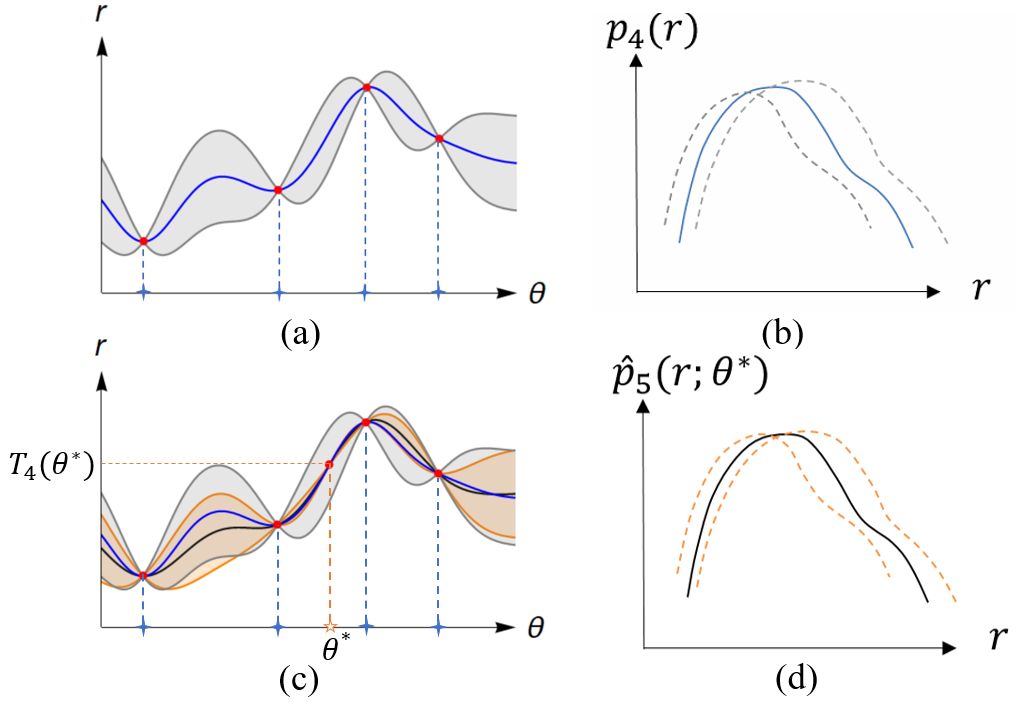}
\captionof{figure}{A schematic plot for the Gaussian process regression (GPR) and the resulted response PDF, by assuming the parameter $\theta$ to be a scalar and $n=4$. (a) predictive mean $T_4(\theta)$ (blue solid line) and uncertainty bounds $T_4(\theta)\pm \sigma_4(\theta)$ (grey solid lines) obtained from four parameter-to-response pairs. (b) $p_4(r)$ (solid line), $p_4^+(r)$ and $p_4^-(r)$ (dashed lines) calculated from $T_4$, $T_4+\sigma_4$ and $T_4-\sigma_4$. (c) predictive mean $\hat{T}_5(\theta;\theta^*)$ (black solid line) and uncertainty bounds $\hat{T}_5(\theta;\theta^*) \pm \hat{\sigma}_5(\theta;\theta^*)$ (orange solid lines) obtained from existing four parameter-to-response pairs and $(\theta^*, T_4(\theta^*))$. (d) $\hat{p}_5(r)$ (solid line), $\hat{p}_5^+(r)$ and $\hat{p}_5^-(r)$ (dashed lines) calculated from $\hat{T}_5$, $\hat{T}_5+\hat{\sigma}_5$ and $\hat{T}_5-\hat{\sigma}_5$. }
    \label{GPRNEW}
\end{center}

We next describe a sequential sampling method which significantly reduces the number of samples (thus computational cost). Compared with the full Monte Carlo (or importance sampling) which generates all samples based only on $p(\bm{\theta})$, our sequential sampling generates samples making use of the previous samples and the corresponding values of $r$ from simulations to stress the tail part. The implementation requires (i) a surrogate model to approximate the map $T$, and (ii) solution of an optimization problem to obtain the next-best sample which provides the fastest convergence rate for the tail part of $p(r)$. The two components are next described in details.

%\begin{center}
%    \includegraphics[width=7.5cm]{GPR14.jpg}
%\captionof{figure}{A schematic plot for the Gaussian process regression (GPR) and the resulted response PDF. (a) predictive mean %$T_n(\theta)$ (blue solid line) and uncertainty bounds $\sigma_n(\theta)$ (grey solid line) obtained from four parameter-to-response %pairs, where the parameter is assumed to be a scalar. (b) $p_n(r)$ (solid line), $p_n^+(r)$ (upper dashed line) and $p_n^-(r)$ (lower %dashed line) calculated from $T_n$, $T_n+\sigma_n$ and $T_n-\sigma_n$.}
%    \label{GPR1}
%\end{center}

For the surrogate model, we use the Gaussian process regression (GPR, a.k.a. Kriging), which is a well-developed method in machine learning \citep{GaussianProcessML} and geostatistics \citep{journel1978mining}. Given a number of available parameter-to-response pairs $\mathcal{D}_n=\{\bm{\theta}^i,r^i=T(\bm{\theta}^i)\}_{i=1}^{i=n}$, the GPR predicts the function $T(\bm{\theta})$ as a realization of a random Gaussian process, whose mean and variance provide, respectively, the approximation of the map $T$ and its uncertainty. We hereafter denote the $n$-pair predictive mean and standard deviation as $T_n(\bm{\theta})$ and $\sigma_n(\bm{\theta})$. The detailed algorithm of GPR based on the Baysian inference is summarized in Appendix B. To visualize the concept, we show a schematic plot in Figure \ref{GPRNEW}(a), where we use $n=4$ and scalar $\theta$ for simplicity. Taking the predictive mean $T_n$ as a (cheap) surrogate model, we are able to calculate the PDF of the response $p_n(r)$ using a large number of samples. In addition, the upper and lower uncertainty bounds of the PDF, $p_n^+(r)$ and $p_n^-(r)$, can be calculated from, say, $T_n+\sigma_n$ and $T_n-\sigma_n$ (Figure \ref{GPRNEW}(b)).

%\begin{center}
%    \includegraphics[width=7.5cm]{GPR24.jpg}
%\captionof{figure}{An illustration of adding one point in GPR. The left figure is the GPR learned from four pairs. The right figure %demonstrates the predictions of adding one more pair, where the output of the fifth pair is generated by the four-pair GPR. The new %predictive values are shown in black line and the uncertainties in orange line. The uncertainties decrease everywhere compared with %four-pair GPR}
%    \label{GPR2}
%\end{center}

Suppose again we have a dataset $ \mathcal{D}_n=\{\bm{\theta}^i,r^i=T(\bm{\theta}^i)\}_{i=1}^{i=n}$ where $T(\bm{\theta}^i)$ is computed by CFD simulations. We aim to find the next-best $(n+1)_{th}$ sample which leads to the fastest convergence of the tail part of $p(r)$. Given an arbitrary next sample $\bm{\theta}^*$, the surrogate model $T_n$ can be used to predict its response $T_n(\bm{\theta}^*)$. This provides us with $n+1$ parameter-to-response pairs $\mathcal{D}_n\cup (\bm{\theta}^*, T_n(\bm{\theta}^*))$, from which a new GRP can be performed to develop an updated surrogate model $\hat{T}_{n+1}(\bm{\theta};\bm{\theta}^*)$, as well as the resulted PDF $\hat{p}_{n+1}(r;\bm{\theta}^*)$. Due to the additional information provided by $(\bm{\theta}^*, T_n(\bm{\theta}^*))$, the uncertainties of both $\hat{T}_{n+1}(\bm{\theta};\bm{\theta}^*)$ and $\hat{p}_{n+1}(r;\bm{\theta}^*)$ are reduced, compared with $T_n(\bm{\theta})$ and $p_n(r)$. The schematic plot of this procedure is shown in Figure \ref{GPRNEW} (c) and (d).

%For this purpose, the approximate map $T_{n-1}$ learned from the available $n-1$ parameter-to-response pairs is used as a surrogate model to predict the response of the $n_{th}$ sample with parameter $\bm{\theta}^*$, i.e., $T_{n-1}(\bm{\theta}^*)$. Then, by using $n$ parameter-to-response pairs $D\cup (\bm{\theta}^*, T_{n-1}(\bm{\theta}^*))$, we can establish a new GPR to predict $\hat{T}_{n}(\bm{\theta};\bm{\theta}^*)$, with reduced uncertainty compared to $T_{n-1}$. The schematic plot of this procedure is shown in figure \ref{GRP2}.

We note that both $\hat{T}_{n+1}(\bm{\theta};\bm{\theta}^*)$ and $\hat{p}_{n+1}(r;\bm{\theta}^*)$ depend on the $(n+1)_{th}$ sample $\bm{\theta}^*$. To find the optimal $\bm{\theta}^*$ for the resolution of the tail of $p(r)$, we construct an optimization problem with an objective function 
\begin{equation}
    Q(\bm{\theta}^*)=\int_0^\infty |\hat{p}_{n+1} ^{+}(r;\bm{\theta}^*)-\hat{p}_{n+1} ^{-}(r;\bm{\theta}^*) | r^sdr
    \label{Qmet}
\end{equation}
where $\hat{p}_{n+1} ^{+}(r;\bm{\theta}^*)$ and $\hat{p}_{n+1} ^{-}(r;\bm{\theta}^*)$ are uncertainty bounds of $\hat{p}_{n+1}(r;\bm{\theta}^*)$ calculated from $\hat{T}_{n+1}+\hat{\sigma}_{n+1}$ and $\hat{T}_{n+1}-\hat{\sigma}_{n+1}$. $r^s$ is a weighting factor which gives more weight for a larger value of $r$ (i.e., tail of $\hat{p}_{n+1} (r;\bm{\theta}^*)$) with $s\gg1$. We use $s=6$ in our current work.

The objective function \eqref{Qmet} is different from the one used in \citet{mohamad2018sequential}, which stresses the low probability part of $\hat{p}_{n+1} (r;\bm{\theta}^*)$ by defining $Q$ based on the difference of log-PDF without the weighting factor. However, a low probability does not necessarily mean a large response (it may also correspond to an extremely small response). In contrast, the function \eqref{Qmet} provides a direct measure of the uncertainty of $\hat{p}_{n+1} (r;\bm{\theta}^*)$ focusing on the part of extreme (large) responses, with the level of ``extreme'' tunable by the value of $s$. We also remark that the sequential sampling based on \eqref{Qmet} can explore all extremes in a multi-modal response function given sufficient samples.

The next-best sample $\bm{\theta}^{n+1}$ is chosen from the sample space by solving the optimization problem
\begin{equation}
    \bm{\theta}^{n+1}={\rm{argmin}}_{\bm{\theta}^*} Q(\bm{\theta}^*).
    \label{opt_Q}
\end{equation}
In our current work, \eqref{opt_Q} is numerically solved using a particle swarm method \citep{poli2007particle}. The sequential sampling process is repeated for the next-best sampling until satisfactory convergence is achieved for the tail part of $p(r)$. We summarize the algorithm of the whole process in the following.

~\\
\noindent{\textbf{Algorithm:}}

\indent{\textbf{input:} initial dataset $\mathcal{D}_{n}=\{\bm{\theta}^i,r^i\}_{i=1}^{i=n}$}

\textbf{repeat:} 

(1) solve \eqref{opt_Q} to find the next-best sample $\bm{\theta}^{n+1}$ according to existing dataset $ \mathcal{D}_n=\{\bm{\theta}^i,r^i\}_{i=1}^{i=n}$. Given $\bm{\theta}^*$, the function $Q(\bm{\theta}^*)$ in \eqref{opt_Q} is calculated by:

\quad I. predict the Gaussian process mean $T_n$ from $\mathcal{D}_n$;

\quad II. predict a new Gaussian process mean $\hat{T}_{n+1}$ and standard deviation $\hat{\sigma}_{n+1}$ based on $ \mathcal{D}_n \cup \{\bm{\theta}^*,T_{n}(\bm{\theta}^*)\}$;

\quad III. generate the upper and lower bounds of the response PDF, $\hat{p}_{n+1} ^{+}(r;\bm{\theta}^*)$ and $\hat{p}_{n+1} ^{-}(r;\bm{\theta}^*)$ from $\hat{T}_{n+1}+\hat{\sigma}_{n+1}$ and $\hat{T}_{n+1}-\hat{\sigma}_{n+1}$;

\quad IV. calculate $Q(\bm{\theta}^*)$;

(2) implement numerical simulation to get the $r^{n+1}(\bm{\theta}^{n+1})$

(3) n=n+1

\textbf{until:} tail of $p_{n+1} (r)$ converges

\textbf{return:} $p_{n+1} (r)$

\section*{FRAMEWORK VALIDATION}

For validation, it is desirable to compare $p(r)$ from our reduced-order approach with the exact motion response PDF $p^e(r)$. This requires a cheap parameter-to-response map $T$ such that $p^e(r)$ can be efficiently calculated. For this purpose, we use a nonlinear roll equation \citep{umeda2004nonlinear} to calculate the time series of ship roll $\xi(t)$:  
%\begin{equation}
\begin{align}
    \ddot{\xi}+\alpha_1 \dot{\xi}+\alpha_2 \dot{\xi}^3 +&(\beta_1+\epsilon_1 \cos (\gamma) \eta(t;\bm{\theta}) )\xi+\beta_2 \xi^3 \nonumber\\
    &= \epsilon_2 \sin (\gamma) \eta(t;\bm{\theta}), 
    \label{roll}
\end{align}
%\end{equation}
which phenomenologically models the ship roll motion due to nonlinear resonance and parametric roll in oblique irregular waves. We use empirical coefficients $\alpha_1=0.25$, $\alpha_2=0.1$, $\beta_1=0.04$, $\beta_2=-0.01$, $\epsilon_1=0.006$, $\epsilon_2=0.008$. Eq.\eqref{roll} is numerically integrated with a 4th-order Runge-Kutta method from zero initial condition to obtain the group-maximum response $r=max(\xi(t))$. The values of parameters are tuned such that the $r$ is not sensitive to the initial conditions for $\eta(t;\bm{\theta})$ described by \eqref{etax} and \eqref{GauWG}.

\begin{figure*}
    \begin{center}
    \includegraphics[width=16cm]{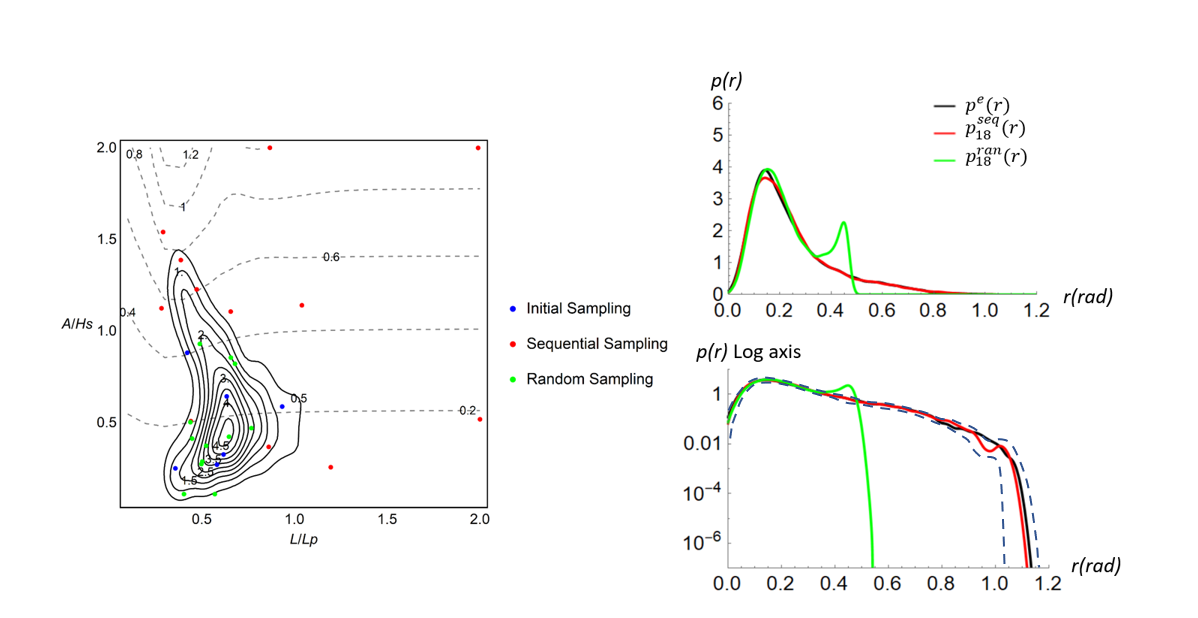}
    \label{12sam}
\captionof{figure}{Samples and PDFs from sequential and random samplings. Left: Initial 6 random samples (blue dots), subsequently 12 sequential samples (red dots) and 12 random samples (green dots) in the parameter space of $(A,L)$. The solid black lines represent $p({A,L})$, and the dashed lines represent $r({A,L})$ (only shown for illustration). Right: The exact roll PDF $p^e(r)$ (black line), sequential-sampling PDF $p_{18}^{seq}(r)$ (red line) and random-sampling PDF $p_{18}^{ran}(r)$ (green line) plotted on both linear and log axes. The 95\% confidence interval for $p_{18}^{seq}(r)$ is included (dashed line) in the log-axis plot.}
\label{12sam}
\end{center}
\end{figure*}

\begin{figure*}
    \begin{center}
    \includegraphics[width=16cm]{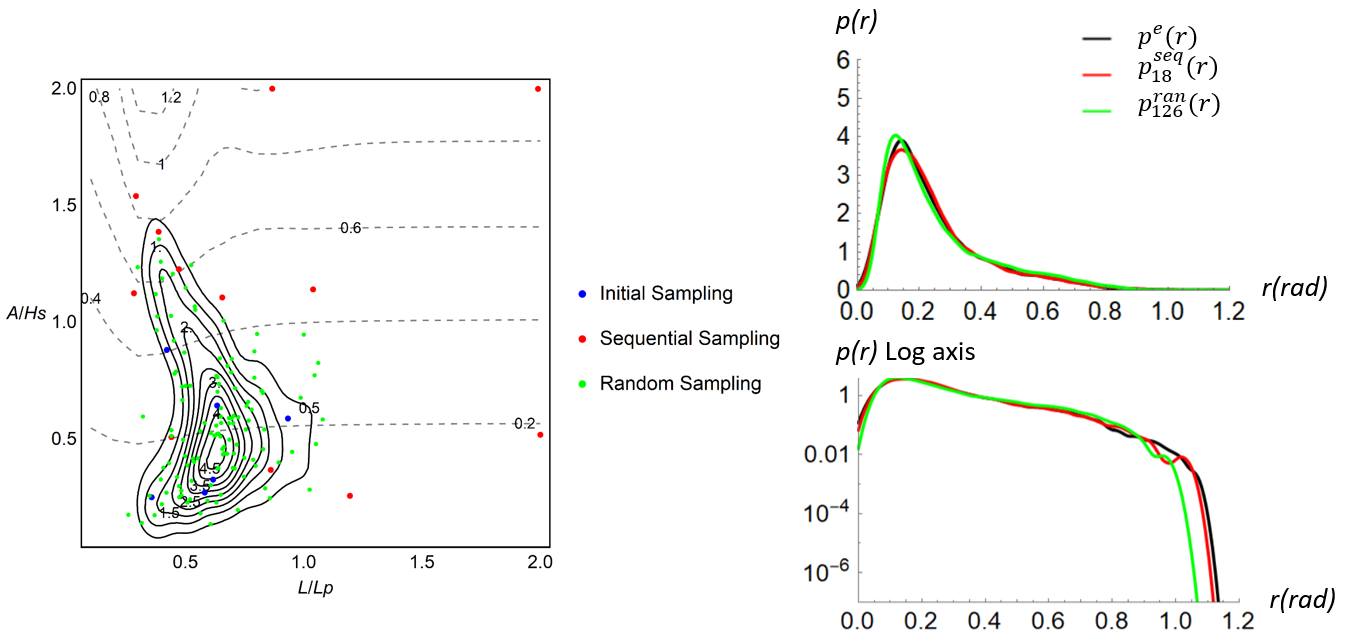}
\captionof{figure}{Samples and PDFs from sequential and random samplings. Left: initial 6 random samples (blue dots), subsequently 12 sequential samples (red dots) and 120 random samples (green dots) in the parameter space of $(A,L)$. The solid black lines represent $p(A,L)$, and the dashed lines represent $r(A,L)$ (only shown for illustration). Right: The exact roll PDF $p^e(r)$ (black line), sequential-sampling PDF $p_{18}^{seq}(r)$ (red line) and random-sampling PDF $p_{126}^{ran}(r)$ (green line) plotted on both linear and log axes.}
\label{120sam}
\end{center}
\end{figure*}

To obtain the time series $\eta(t;\bm{\theta})$ for a given $\bm{\theta}$, we first construct the localized Gaussian group from \eqref{GauWG} and then the spatial surface elevation $\eta(x;\bm{\theta})$ from \eqref{etax}. The effect of $\phi(x)$ is neglected in \eqref{etax}, since the phase modulation $\phi(x)$ can be considered almost a constant within a wave group (see Figure \ref{HT}(b)) for a sufficiently narrow-band wave spectrum. The relatively large variation of $\phi(x)$ occurs near the end of the wave group, which is expected to have a much smaller impact on the extreme motion response than $\bm{\theta}\equiv(A,L)$. Given $\eta(x;\bm{\theta})$, $\eta(t;\bm{\theta})$ is then constructed by considering the propagation of each wave mode by linear dispersion velocity.

To calculate $p^e(r)$, we generate $\rho_c(t)$ from $\rho_c(x)$ (Figure \ref{HT}(a)) in a similar manner, which is then substituted to \eqref{roll} in replacement of $\eta(t;\bm{\theta})$. This is conducted for all $\rho_c(x)$ in the ensemble of nonlinear wave fields to obtain a collection of $r$, which is then used to calculate $p^e(r)$.

We next describe the implementation of our framework on this simplified ship roll problem. We consider an initial narrow-band wave spectrum in a Gaussian form:
\begin{equation}
F(k)\sim \exp \frac{-(k-k_0)^2}{2\mathcal{K}^2},  
\label{Fk}
\end{equation}
with significant wave height $H_s=12m$, peak (carrier) wavenumber $k_0=0.018 m^{-1}$ (corresponding to peak period $T_p=15s$), and $\mathcal{K}=0.05k_0$. Four hundred HOS simulations are run, where each of them has a domain length of 64 times the peak wavelength. The ensemble of nonlinear wave fields are collected at $t=50T_p$, for which the wave parameterization method discussed before is used to generate the joint PDF of $A$ and $L$.

%\begin{figure}[H]
%\centering
%\subfigure[]{
%\label{Fig.sub.1}
%\includegraphics[width=7cm]{C22.pdf}}
%\subfigure[]{
%\label{Fig.sub.2}
%\includegraphics[width=7CM]{C42.pdf}}
%\caption{Comparison of exact PDF, PDF generated by 12 sequential samplings, 120 sequential samples and random samplings: (a) linear axis (b) Log axis.}
%\label{Fig.lable}
%\end{figure}

The sampling process in the space $(A,L)$ is performed as follows. We first generate six random samples (following $p(A,L)$) and calculate their response $r$ from \eqref{roll} as the initial dataset. Then we conduct twelve sequential samplings to compute $p_{18}^{seq}(r)$. For comparison, a random sampling approach with an equal number of samples is also conducted to compute $p_{18}^{ran}(r)$ (based on the GPR surrogate model constructed from 18 random samples).

The PDFs $p_{18}^{seq}(r)$, $p_{18}^{ran}(r)$ and $p^e(r)$, as well as the 95\% confidence interval for $p_{18}^{seq}(r)$, are plotted in Figure \ref{12sam}, on both linear and log scales. It can be seen that the result from the sequential sampling is much closer to the exact PDF compared to the result from the random sampling. With the left sub-figure showing the locations of the samples, we see that the random samples are concentrated in the high probability region of $p(A,L)$, while the sequential samples explore the region with large response (combined with nontrivial probability). In order to obtain comparable roll PDF using random sampling, at least one order of magnitude more samples are needed. Figure \ref{120sam} shows the result $p_{126}^{ran}(r)$ obtained from 120 random samples (after the 6 initial ones), which is still less accurate than $p_{18}^{seq}(r)$ in terms of the tail part of the PDF.   

We have now demonstrated the effectiveness of both wave parameterization and sequential sampling. This allows us to couple our framework with a more advanced CFD tool to compute the extreme ship motion statistics in the next section.

\section*{EFFECTS OF WAVE NONLINEARITY}
Our framework allows the direct resolution of wave nonlinearity effect on the ship response statistics. It has been experimentally and numerically demonstrated that the modulational instability of nonlinear waves result in the non-Gaussian statistics (in terms of a heavy-tail PDF) of the surface elevations \citep{onorato2009statistical,xiao2013rogue}. Based on this, we expect the extreme ship response statistics to be enhanced when nonlinearity effect is included for the wave field.

\begin{figure*}
\begin{subfigure}
\centering
  \subfigure{\includegraphics[trim = 0mm 0mm 0mm 0mm, clip, keepaspectratio, width=14cm]{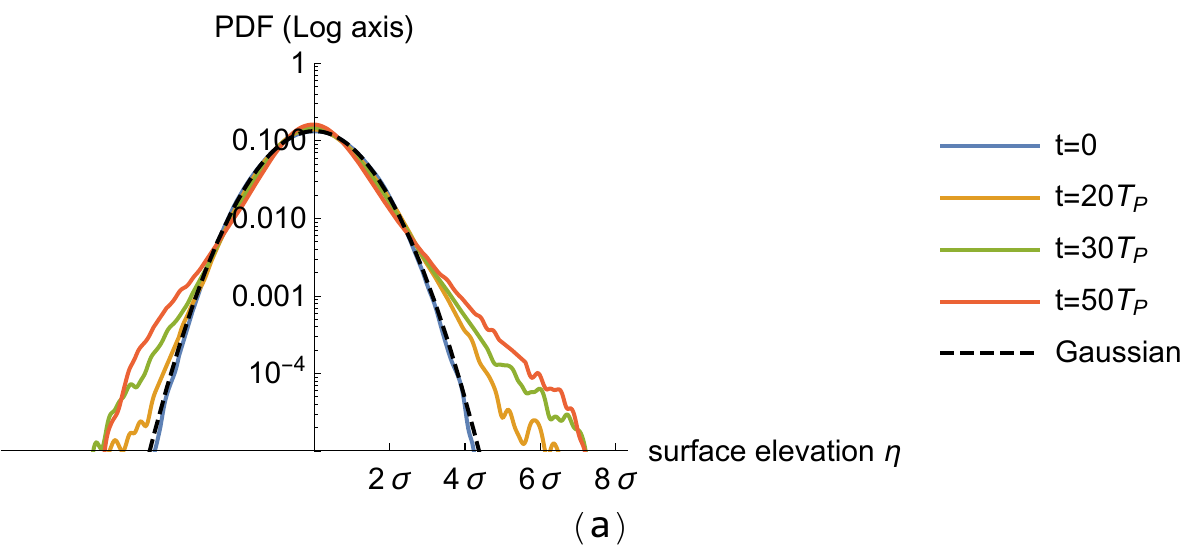}}
  \label{fig:1a}
  \subfigure{\includegraphics[trim = 0mm 0mm 0mm 0mm, clip, keepaspectratio, width=14cm]{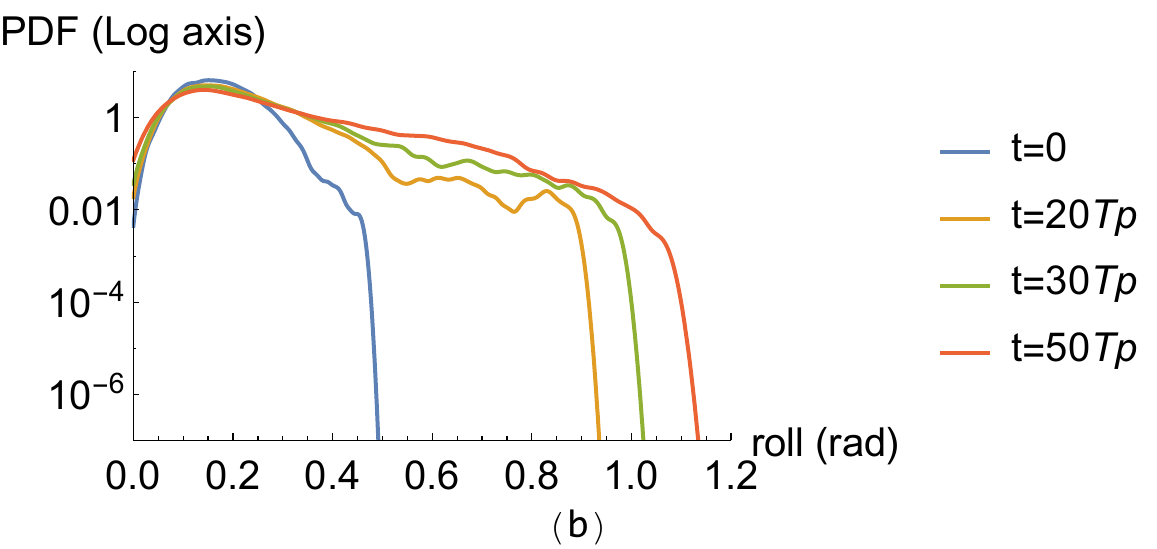}}
  \label{fig:1b}
 \caption{PDFs of (a) surface elevation and (b) roll response at t=0 (blue), 20$T_p$ (orange), 30$T_p$ (green) and 50$T_p$ (red). The Gaussian PDF is plotted in (a) with a black dashed line.}
 \label{fig:nonlinear}
\end{subfigure}
\end{figure*}

To illustrate the effect of wave nonlinearity, we generate an initial linear wave field from the spectrum \eqref{GauWG} using independent and random phase distributions, i.e., with surface elevation following Gaussian statistics. An ensemble of four hundred HOS simulations (with domain length of 64 times of the peak wavelength) are used to compute the evolution of wave statistics with time. Figure \ref{fig:nonlinear}(a) shows the PDFs of surface elevations at $t=0$, $20T_p$, $30T_p$ and $50T_p$ respectively. It is clear that the PDF at $t=0$ follows closely a Gaussian tail. With the increase of time, a heavier tail develops due to the nonlinear wave effect, indicating a higher probability of extreme waves in the field. We compute the response PDF using roll equation and sequential sampling with wave group statistics calculated from the wave fields in each instant of figure \ref{fig:nonlinear}(a), with the results shown in figure \ref{fig:nonlinear}(b). We can clearly observe the development of enhanced extreme ship motion statistics with the increase of time. Since the evolution of wave spectrum in $O(10-50)T_p$ is relatively mild, it is not unreasonable to associate the enhanced motion statistics to non-Gaussian wave statistics developed due to the wave nonlinearity. While this case qualitatively demonstrates the influence of wave nonlinearity on extreme ship motion statistics, more studies are warrantied in future work to quantify this effect.

\section*{COUPLING WITH CFD}
In this section, we use CFD simulations to compute the ship motion response in a given incident wave group described by parameters $A$ and $L$. For each sample in the wave parameter space, we define the initial condition of CFD simulation using the profile of a propagating wave group with parameters $A$ and $L$, and compute the maximum response as the output of the simulation. For simplicity, we consider the motion of a two-dimensional (2D),  square-shaped hull geometry with $40m\times40m$ cross section and density $\rho_h=0.5\rho_w$ with $\rho_w$ being the water density. The turbulence effects are neglected in the simulations. Our interest is to resolve the extreme roll statistics using the framework described above. In spite of the simplification (in terms of the 2D geometry), this computation is sufficient to demonstrate the effectiveness of the new framework when coupled to CFD, which enables more realistic problems to be studied. 

\subsection*{CFD model}
The CFD simulations in this work are performed using the open-source code OpenFOAM \citep{Jasak_2009}. We next describe the details of our model. 

\subsubsection*{Mathematical formulation}
The air-water interface in CFD is modeled by the volume fraction $\gamma$ ($ \gamma=0 $ for air and $ \gamma = 1 $ for water). To capture the evolution of the $\gamma$ field, we use the interFoam solver, which is based on an algebraic volume of fluid (AVOF) approach. In AVOF, the flux of volume fraction $\gamma$ is computed algebraically without a geometric reconstruction of the interface \citep{Mirjalili_2017}. An interfacial compression term is used to mitigate the effects of numerical smearing of the interface \citep{Deshpande_2012}. The governing equations of the problem include the continuity equation, the momentum equation and the volume fraction equation:
\begin{equation}
    \nabla\cdot\mathbf{u} = 0,
\end{equation}
\begin{equation}
\begin{aligned}
    \frac{\partial(\rho\mathbf{u})}{\partial t} + \nabla\cdot(\rho\mathbf{uu}) =& -\nabla p_d + \nabla\cdot(\mu\nabla\mathbf{u}) \\ 
    & + \nabla\mathbf{u}\cdot\nabla\mu - \mathbf{g}\cdot\mathbf{x}\nabla\rho,
\end{aligned}
\end{equation}
\begin{equation}
    \frac{\partial\gamma}{\partial t} + \nabla\cdot(\mathbf{u}\gamma) + \nabla\cdot(\mathbf{u}_r\gamma(1-\gamma))=0,
    \label{gamma}
\end{equation}
where $ \mathbf{u} $ is the velocity field, $ p_d = p - \rho\mathbf{g}\cdot\mathbf{x} $ is the modified pressure, with $p$ being the pressure, and $ \mathbf{g} $ the gravitational acceleration vector, and $ \mathbf{x} $ the position vector. The fluid properties, including the density $ \rho $ and the dynamic viscosity $ \mu $ are calculated as weighted averages based on $\gamma$,
\begin{align*}
    \rho &= \gamma\rho_w + (1-\gamma)\rho_a, \\
    \mu &=  \gamma\mu_w + (1-\gamma)\mu_a,
\end{align*}
where the subscripts $ w $ and $ a $ correspond to water and air respectively. In \eqref{gamma}, the last term (with tuned parameter $ \mathbf{u}_r $) represents an artificial compression of the interface to mitigate the numerical diffusion \citep{Berberovi_2009}. 

The motion of the hull is calculated based on the force exerted by flow pressure and shear stress. The hull is considered as a rigid body, whose motion is solved by numerical integration implemented by the Newmark method \citep{Newmark_1959}:
\begin{equation}
    \dot{X}_{n+1} = \dot{X}_n + (1-\gamma)\Delta t\ddot{X}_n + \gamma\Delta t\ddot{X}_{n+1},
\end{equation}
\begin{equation}
    X_{n+1} = X_n + \Delta t\dot{X}_n + \frac{1-2\beta}{2}\Delta t^2\ddot{X}_n + \beta\Delta t^2\ddot{X}_{n+1},
\end{equation}
where $ X $ represents the linear displacement of the heave and sway motion, and the angular displacement of the roll motion, $ \Delta t $ is the time step, $ \gamma $ and $ \beta $ are user-defined coefficients. Typically, $ \gamma = 0.5, \beta = 0.25 $ are the most common choices and are used in our study.

\subsubsection*{Computational grids}
The 2D computational domain is discretized by a Cartesian mesh with refined grids near the free surface (see Figure \ref{grid}(a)). The dynamic mesh is used around the hull, allowing the mesh to deform when the hull moves. The region of mesh deformation is controlled both inner and outer distance: the grids within the inner distance from the body surface move with the hull as a rigid body; the grids between the inner distance and the outer distance can be morphed; and the grids outside the outer distance do not move (see Figure \ref{grid}(b)).

\subsubsection*{Initial condition and boundary conditions}
The initial condition of the simulation includes the initial fields $\gamma_0(x,z) \equiv \gamma(x,z,t=0)$ and $\mathbf{u}_0(x,z) \equiv \mathbf{u}(x,z,t=0)$, where $z$ is the coordinate in the vertical direction and $t$ is the time. These initial fields are defined separately for the left and right parts of the computation domain (see Figure \ref{window}). For the right part, we consider the situation of a stationary hull floating on still water, with a corresponding volume fraction $\gamma_0(x,z)$, and velocity $\mathbf{u}_0(x,z)=\mathbf{0}$. For the left part, the initial condition represents the profile of a propagating wave group with parameters $A$ and $L$. In particular, the $\gamma_0(x,z)$ field is specified from the free surface position $\eta_0(x) = \eta(x; A,L)$ (see \eqref{etax} and \eqref{GauWG}) by
\begin{equation}
    \gamma_0(x,z) = \left\{
    \begin{aligned}
    &0, & z > \eta_0(x) \\
    &1, & z < \eta_0(x)
    \end{aligned}
    \right.
    \label{etatogamma}
\end{equation}
The $\mathbf{u}_0(x,z)$ field is calculated correspondingly assuming a linear dispersion relation for each wave mode of $\eta_0(x)$, namely with the horizontal and vertical velocity components
\begin{equation}
    u_0(x,z) = \sum_{j=1}^N a_j\omega_j\frac{\cosh[k_j(z+h)]}{\sinh(k_jh)}\cos(k_jx+\psi_j),
\end{equation}
\begin{equation}
    v_0(x,z) = \sum_{j=1}^N a_j\omega_j\frac{\sinh[k_j(z+h)]}{\sinh(k_jh)}\sin(k_jx+\psi_j),
\end{equation}
where $a_j\cos(k_jx+\psi_j)$ is the mode $j$ of $\eta_0(x)$, with $a_j$ the amplitude, $k_j$ the modal wavenumber, and $\psi_j$ the modal phase. $ \omega_j = \sqrt{gk_j\tanh(k_jh)} $ is the angular frequency of mode $j$, $ h $ is the water depth (300m in our case).

\begin{center}
\includegraphics[width=7cm]{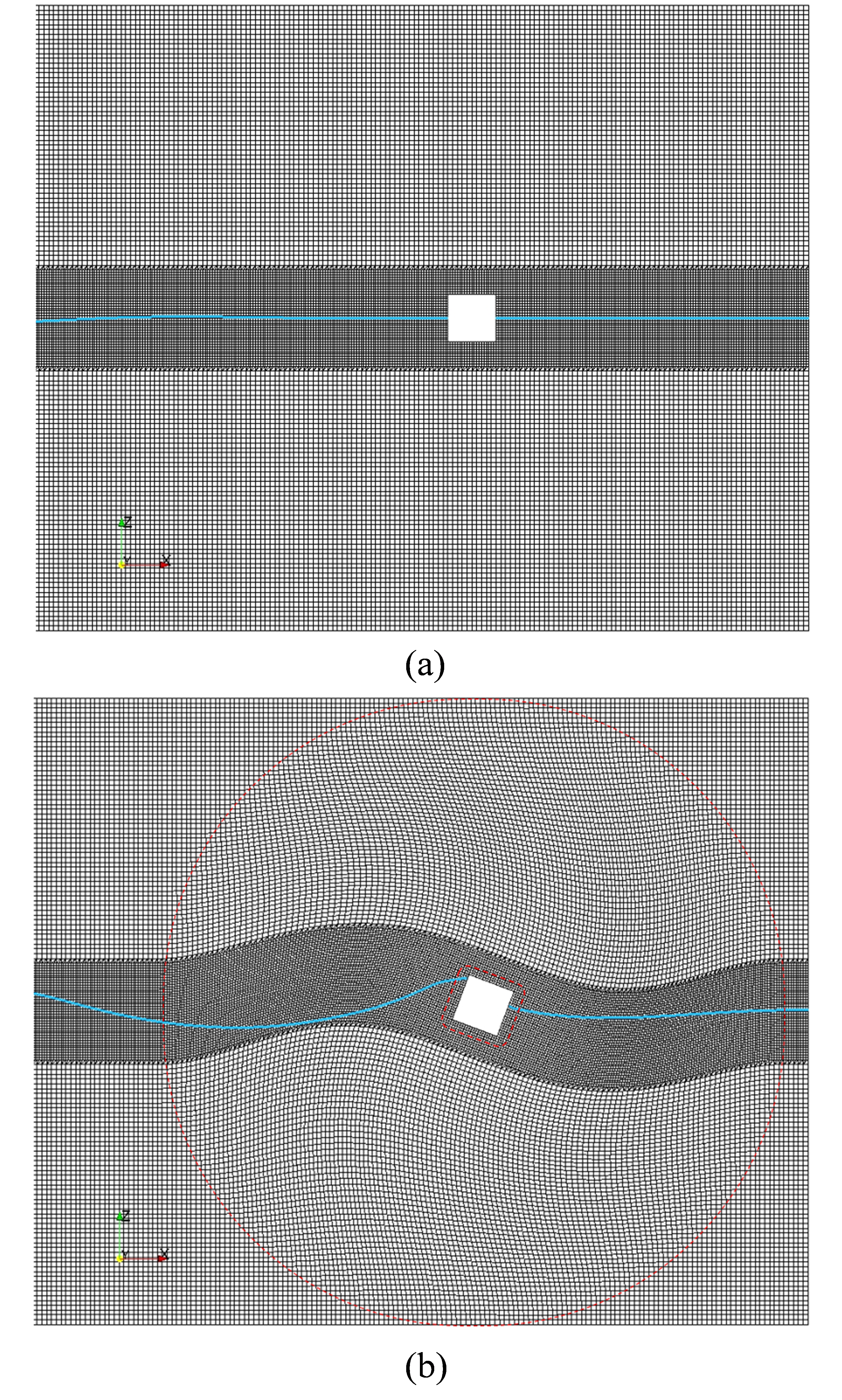}
\captionof{figure}{(a) The mesh around the hull with a refined region near the free surface; (b) The deformed mesh around the moving hull. The interface is marked by a blue solid line. The boundaries of inner distance and outer distance are marked by red dashed lines.}
\label{grid}
\end{center}
\begin{center}
    \includegraphics[width=8cm]{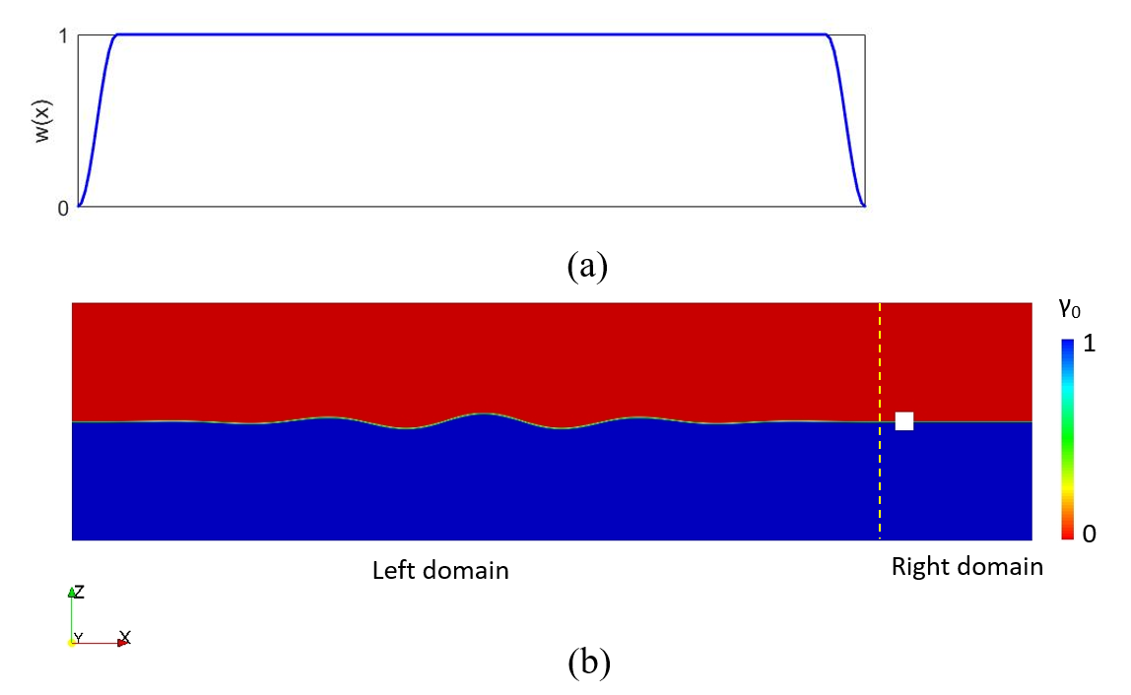}
\captionof{figure}{(a) the window function $w(x)$; (b) a typical initial field $ \gamma_0(x,z) $. The border of the left and right domains is indicated by a dashed line.}
\label{window}
\end{center}

To obtain a smooth field as the initial condition, a window function $w(x)$ is multiplied to $ \gamma_0(x,z) $ and $ \mathbf{u}_0(x,z) $ to remove the discontinuity between the wave field and the still-water field at both ends of the wave group. This is illustrated in Figure \ref{window} along with the resulted initial $ \gamma_0(x,z) $ field.   

Periodic boundary conditions are used at the left and right ends of the domain, allowing the wave to keep propagating without reflection after interacting with the hull. We have tested that the domain is sufficiently large such that the periodic boundary does not result in spurious waves interacting with the body (in the time period of the simulation). At the bottom of the domain, a slip-wall boundary condition is used. At the top, the pressureInletOutletlVelocity boundary condition is used, which is a modified zero-gradient velocity condition\footnote{\url{https://www.openfoam.com/documentation/guides/latest/api/classFoam_1_1pressureInletOutletVelocityFvPatchVectorField.html}}. For the boundaries on the floating hull, the no-slip boundary condition is used, which sets the fluid velocity to the same value as the velocity of the moving rigid body.

\subsubsection*{A typical case}
We show the result of a typical case with $A=8.1m$ and $L=332.7m$ in Figure \ref{CFD}. A snapshot of the $\gamma(x,z)$ field is shown in Figure \ref{CFD}(a) in the process of a wave group interacting with the hull. The time series of roll angle $\xi(t)$ is shown in Figure \ref{CFD}(b). The group-maximum response is then taken as $r=max(\xi(t))$, which is used in the sequential sampling.

\subsection*{Results}
The wave parameter space $p(\bm{\theta})$ is generated using the initial condition \eqref{Fk} with data collected at $t=20T_p$. The sequential sampling is coupled with CFD simulations to compute $p(r)$. We again use 6 random samples (as well as $r(\bm{\theta})$ obtained from CFD) as the initial dataset. Then 14 sequential samples are performed to compute $p_{20}(r)$. Since the exact $p(r)$ is not available for this problem, we plot $p_n(r)$ for different $n$ in Figure \ref{PDFPlot}. We see that there is a clear trend of convergence for the tail of $p_n(r)$ as $n$ increases. 
%In practice, the sampling can be stopped when the difference between $p_n(r)$ and $p_{n+1}(r)$ is sufficiently small.
For the last few samples ($n=16\sim20$), the tail of the PDF oscillates in small regions, indicating the convergence of the extreme statistics.

\begin{center}
    \includegraphics[width=7.5cm]{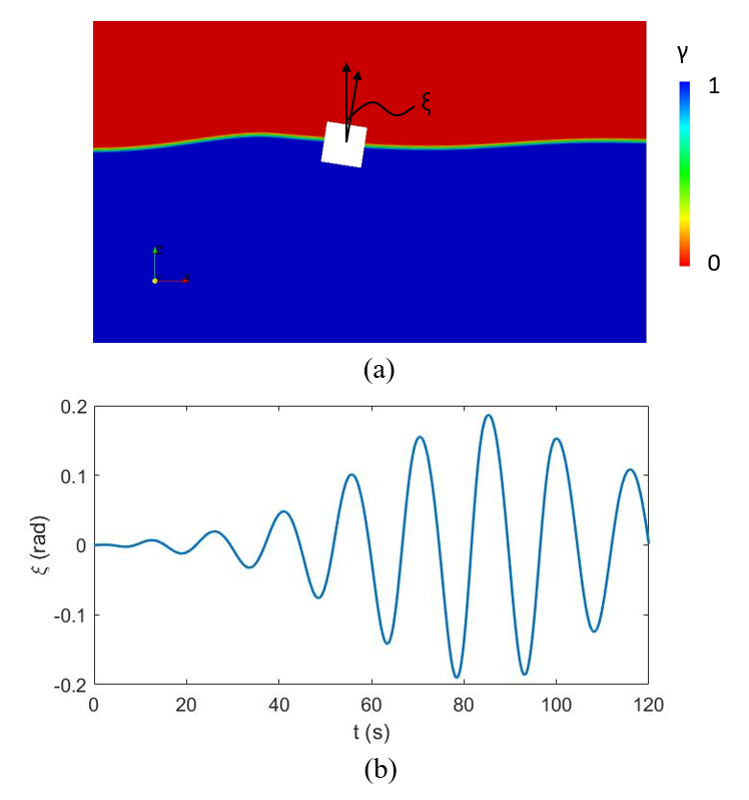}
\captionof{figure}{The result from a typical case with $A=8.1m$ and $L=332.7m$. (a) a snapshot of the $\gamma(x,z)$ field in the process of a wave group interacting with the hull; (b) time series of $\xi(t)$.}
\label{CFD}
\end{center}

\begin{center}
    \includegraphics[width=8cm]{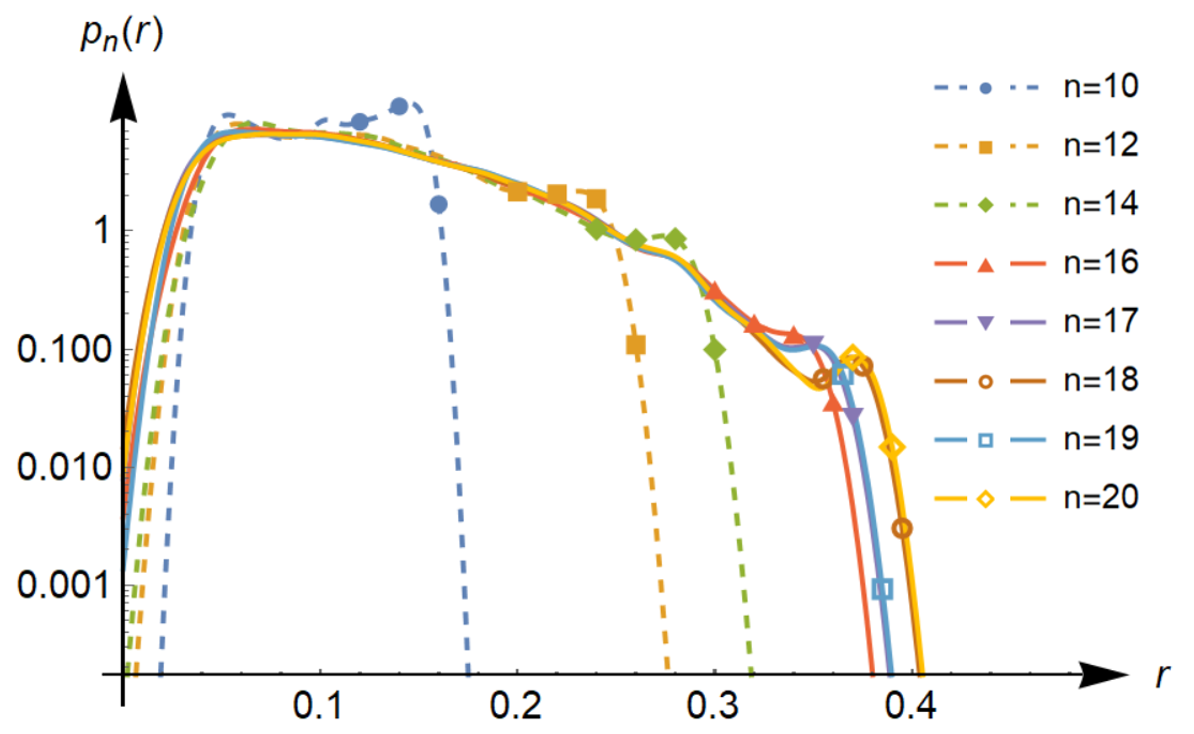}
\captionof{figure}{The PDFs $p_n(r)$ for different $n$.}
\label{PDFPlot}
\end{center}

\begin{center}
    \includegraphics[width=8cm]{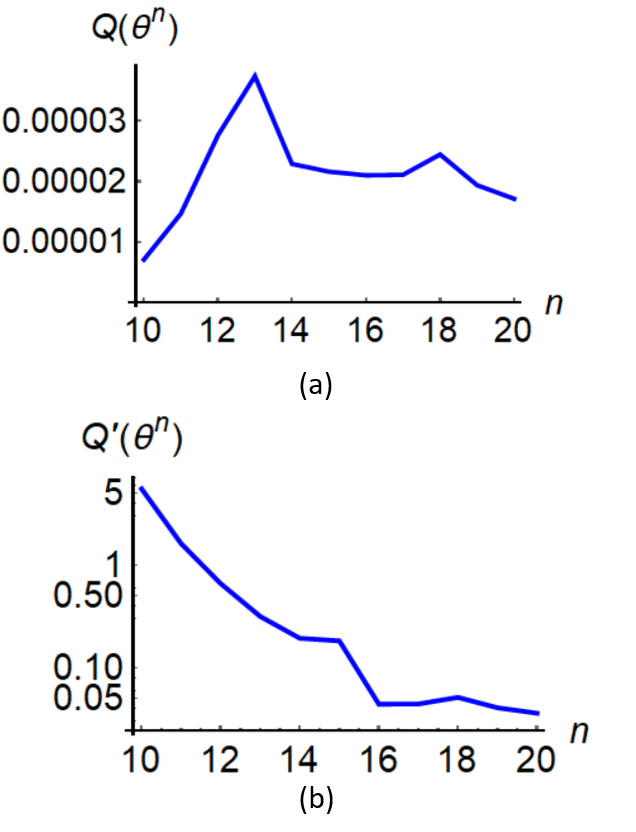}
\captionof{figure}{ (a) $Q(\bm{\theta}^n)$ as a function of the sample number $n$; (b) $Q'(\bm{\theta}^n)$ as a function of the sample number $n$}
\label{QPlot}
\end{center}
We also plot $Q(\bm{\theta}^n)$ as a function of the sample number $n$ in Figure \ref{QPlot}(a). We see that at the end of 20 samples, the value of $Q(\bm{\theta}^n)$  almost converges to a constant level. Nevertheless, this constant level is higher than $Q$ at previous samples, e.g., $Q(\bm{\theta}^{20})>Q(\bm{\theta}^{10})$. This is due to the shift of the tail of $p_n(r)$ to the right (or larger values of $r$), which results in an increase of $Q$ through the factor $r^s$ (see \eqref{Qmet}). An alternative metric to quantify the uncertainty level of the PDF tail can be defined as $Q'=Q(\bm{\theta}^n)/[max(T_n(\bm{\theta}))]^7$, where $max(T_n(\bm{\theta}))$ provides the maximum value of response from all parameters $\bm{\theta}$. Compared to $Q$, $Q'$ removes the effect of the right-shifting of the PDF tail, and focuses only on the difference between the upper and lower bounds of the tail. 
The metric $Q'(\bm{\theta}^n)$ as a function of $n$ is plotted in Figure \ref{QPlot}(b), showing a satisfactory convergence over all samples.

%figure yy caption: 
%figure zz caption: (a) $Q(\bm{\theta}^n)$ as a function of the sample number $n$; (b) $Q'(\bm{\theta}^n)$ as a function of the sample number $n$.

%Now we couple the CFD simualtion with our framework to generate the extreme response statistics. The same wave field is used as the wave input in this CFD coupled framework. Then we conduct our sampling strategy, generating 15 sequential samples one by one after 6 random samples. The PDF of roll angel is shown in figure. We also demonstrate the variation of Q (uncertainty measure, eq. (4)). Initially, Q increases with the evolving of PDF towards to large response direction as our sequential sampling method keeps draw samples to explore larger response area. After certain sequential samples, the range of PDF function with nonzero value become stable. Further, samplings will focus on reduce the uncertainty in this range. 

\section*{CONCLUSIONS AND DISCUSSION}
Building on existing methods, we implement a computational framework which allows an efficient resolution of extreme ship motion statistics in narrow-band nonlinear wave fields. Three key components are included in the framework: (1) generation of an ensemble of nonlinear wave fields using the high-order spectral method; (2) wave group parameterization to reduce the high-dimensional wave field to a low-dimensional space of $(A,L)$; and (3) sequential sampling to obtain the motion response PDF $p(r)$ with fastest convergence rate of its tail, i.e., extreme motion part. In addition to some improvements to the existing methods, our framework allows the effect of wave nonlinearity to be incorporated in the computation of ship response statistics. The framework is validated through a simplified problem of roll motion predicted by a nonlinear roll equation, where the sequential sampling are shown to be effective in obtaining accurate $p(r)$ with significantly reduced computational cost. The capability of the framework to include nonlinearity of the waves have also been demonstrated in an evolving nonlinear wave field. We finally demonstrate the coupling of the framework with CFD to resolve the extreme roll statistics of a two-dimensional, square-shaped hull. 

We note that three approximations in our method require further considerations: (1) The Gaussian wave group representation neglects the deviation of realistic wave groups from Gaussian functions; (2) The application of a constant phase modulation ignores its variation within a wave group; (3) The use of zero initial conditions neglects the high variability of initial conditions in a wave field with multiple groups. The impact of factors (1) and (2) can be considered insignificant in the limit of narrow-band wave field, with their effects increasing with the increase of spectral bandwidth. The impact of (3) is negligible only for dynamical systems where the maximum responses are not sensitive to initial conditions (such as the roll equation considered in this paper). For general cases, the uncertainties associated with (1)-(3) have to be considered. We are now developing an improved framework to account for all these uncertainties. In addition, the quantification of effect of wave nonlinearity on extreme motion statistics will be another direction in our future work.

\section*{ACKNOWLEDGEMENT}
This research is supported by Office of Naval Research grant N00014-20-1-2096. We thank the program manager Dr. Woei-Min Lin for several helpful discussions on the research, and Dr. Venkat Raman for helpful suggestions to improve the paper. This work used the Extreme Science and Engineering Discovery Environment (XSEDE) [Towns 2014] through allocation TG-BCS190007.

\bibliographystyle{plainnat}
\bibliography{mybib.bib}

\begin{thebibliography}{45}
\providecommand{\natexlab}[1]{#1}
\providecommand{\url}[1]{\texttt{#1}}
\expandafter\ifx\csname urlstyle\endcsname\relax
  \providecommand{\doi}[1]{doi: #1}\else
  \providecommand{\doi}{doi: \begingroup \urlstyle{rm}\Url}\fi

\bibitem[Anastopoulos and Spyrou(2016)]{anastopoulos2016ship}
Panayiotis~A Anastopoulos and Kostas~J Spyrou.
\newblock Ship dynamic stability assessment based on realistic wave group
  excitations.
\newblock \emph{Ocean Engineering}, 120:\penalty0 256--263, 2016.

\bibitem[Anastopoulos and Spyrou(2019)]{anastopoulos2019evaluation}
Panayiotis~A Anastopoulos and Kostas~J Spyrou.
\newblock Evaluation of the critical wave groups method in calculating the
  probability of ship capsize in beam seas.
\newblock \emph{Ocean Engineering}, 187:\penalty0 106213, 2019.

\bibitem[Anastopoulos et~al.(2016)Anastopoulos, Spyrou, Bassler, and
  Belenky]{anastopoulos2016towards}
Panayiotis~A Anastopoulos, Kostas~J Spyrou, Christopher~C Bassler, and Vadim
  Belenky.
\newblock Towards an improved critical wave groups method for the probabilistic
  assessment of large ship motions in irregular seas.
\newblock \emph{Probabilistic Engineering Mechanics}, 44:\penalty0 18--27,
  2016.

\bibitem[Bassler et~al.(2010{\natexlab{a}})Bassler, Belenky, and
  Dipper]{bassler2010application}
Christopher~C Bassler, Vadim Belenky, and MJ~Dipper.
\newblock Application of wave groups to assess ship response in irregular seas.
\newblock In \emph{Proc. 11th int. ship stability workshop},
  2010{\natexlab{a}}.

\bibitem[Bassler et~al.(2010{\natexlab{b}})Bassler, Belenky, and
  Dipper~Jr]{bassler2010characteristics}
Christopher~C Bassler, Vadim Belenky, and Martin~J Dipper~Jr.
\newblock Characteristics of wave groups for the evaluation of ship response in
  irregular seas.
\newblock In \emph{International Conference on Offshore Mechanics and Arctic
  Engineering}, volume 49101, pages 227--237, 2010{\natexlab{b}}.

\bibitem[Bassler et~al.(2019)Bassler, Dipper, and
  Melendez]{bassler2019experimental}
Christopher~C Bassler, Martin~J Dipper, and Mark Melendez.
\newblock Experimental ship dynamic stability assessment using wave groups.
\newblock In \emph{Contemporary Ideas on Ship Stability}, pages 507--520.
  Springer, 2019.

\bibitem[Belenky et~al.(2018)Belenky, Weems, Pipiras, Glotzer, and
  Sapsis]{belenky2018tail}
V~Belenky, K~Weems, V~Pipiras, D~Glotzer, and T~Sapsis.
\newblock Tail structure of roll and metric of capsizing in irregular waves.
\newblock In \emph{Proc. 32nd Symp. Naval Hydrodynamics, Hamburg, Germany},
  2018.

\bibitem[Belenky et~al.(2012)Belenky, Weems, Bassler, Dipper, Campbell, and
  Spyrou]{belenky2012approaches}
Vadim Belenky, Kenneth~M Weems, Christopher~C Bassler, Martin~J Dipper,
  Bradley~L Campbell, and Kostas~J Spyrou.
\newblock Approaches to rare events in stochastic dynamics of ships.
\newblock \emph{Probabilistic Engineering Mechanics}, 28:\penalty0 30--38,
  2012.

\bibitem[Belenky et~al.(2016)Belenky, Weems, and Lin]{belenky2016split}
Vadim Belenky, Kenneth Weems, and Woei-Min Lin.
\newblock Split-time method for estimation of probability of capsizing caused
  by pure loss of stability.
\newblock \emph{Ocean Engineering}, 122:\penalty0 333--343, 2016.

\bibitem[Berberovi\ifmmode~\acute{c}\else \'{c}\fi{}
  et~al.(2009)Berberovi\ifmmode~\acute{c}\else \'{c}\fi{}, van Hinsberg,
  Jakirli\ifmmode~\acute{c}\else \'{c}\fi{}, Roisman, and
  Tropea]{Berberovi_2009}
Edin Berberovi\ifmmode~\acute{c}\else \'{c}\fi{}, Nils~P. van Hinsberg, Suad
  Jakirli\ifmmode~\acute{c}\else \'{c}\fi{}, Ilia~V. Roisman, and Cameron
  Tropea.
\newblock Drop impact onto a liquid layer of finite thickness: Dynamics of the
  cavity evolution.
\newblock \emph{Phys. Rev. E}, 79:\penalty0 036306, 2009.

\bibitem[Boccotti(1997)]{boccotti1997general}
Paolo Boccotti.
\newblock A general theory of three-dimensional wave groups part i: the formal
  derivation.
\newblock \emph{Ocean engineering}, 24\penalty0 (3):\penalty0 265--280, 1997.

\bibitem[Boccotti(2008)]{boccotti2008quasideterminism}
Paolo Boccotti.
\newblock Quasideterminism theory of sea waves.
\newblock \emph{Journal of offshore mechanics and Arctic engineering},
  130\penalty0 (4), 2008.

\bibitem[Campbell et~al.(2016)Campbell, Belenky, and
  Pipiras]{campbell2016application}
Bradley Campbell, Vadim Belenky, and Vladas Pipiras.
\newblock Application of the envelope peaks over threshold (epot) method for
  probabilistic assessment of dynamic stability.
\newblock \emph{Ocean Engineering}, 120:\penalty0 298--304, 2016.

\bibitem[Cousins and Sapsis(2016)]{cousins2016reduced}
Will Cousins and Themistoklis~P Sapsis.
\newblock Reduced-order precursors of rare events in unidirectional nonlinear
  water waves.
\newblock \emph{Journal of Fluid Mechanics}, 790:\penalty0 368--388, 2016.

\bibitem[Cousins et~al.(2019)Cousins, Onorato, Chabchoub, and
  Sapsis]{cousins2019predicting}
Will Cousins, Miguel Onorato, Amin Chabchoub, and Themistoklis~P Sapsis.
\newblock Predicting ocean rogue waves from point measurements: An experimental
  study for unidirectional waves.
\newblock \emph{Physical Review E}, 99\penalty0 (3):\penalty0 032201, 2019.

\bibitem[Dematteis et~al.(2019)Dematteis, Grafke, Onorato, and
  Vanden-Eijnden]{dematteis2019experimental}
Giovanni Dematteis, Tobias Grafke, Miguel Onorato, and Eric Vanden-Eijnden.
\newblock Experimental evidence of hydrodynamic instantons: the universal route
  to rogue waves.
\newblock \emph{Physical Review X}, 9\penalty0 (4):\penalty0 041057, 2019.

\bibitem[Deshpande et~al.(2012)Deshpande, Anumolu, and
  Trujillo]{Deshpande_2012}
Suraj~S Deshpande, Lakshman Anumolu, and Mario~F Trujillo.
\newblock Evaluating the performance of the two-phase flow solver {interFoam}.
\newblock \emph{Computational Science {\&} Discovery}, 5\penalty0 (1):\penalty0
  014016, 2012.

\bibitem[Dommermuth and Yue(1987)]{dommermuth1987high}
Douglas~G Dommermuth and Dick~KP Yue.
\newblock A high-order spectral method for the study of nonlinear gravity
  waves.
\newblock \emph{Journal of Fluid Mechanics}, 184:\penalty0 267--288, 1987.

\bibitem[Echard et~al.(2011)Echard, Gayton, and Lemaire]{echard2011ak}
B~Echard, N~Gayton, and M~Lemaire.
\newblock Ak-mcs: an active learning reliability method combining kriging and
  monte carlo simulation.
\newblock \emph{Structural Safety}, 33\penalty0 (2):\penalty0 145--154, 2011.

\bibitem[Hu and Mahadevan(2016)]{hu2016global}
Zhen Hu and Sankaran Mahadevan.
\newblock Global sensitivity analysis-enhanced surrogate (gsas) modeling for
  reliability analysis.
\newblock \emph{Structural and Multidisciplinary Optimization}, 53\penalty0
  (3):\penalty0 501--521, 2016.

\bibitem[Jasak(2009)]{Jasak_2009}
Hrvoje Jasak.
\newblock Open{FOAM}: Open source {CFD} in research and industry.
\newblock \emph{International Journal of Naval Architecture and Ocean
  Engineering}, 1\penalty0 (2):\penalty0 89–94, 2009.

\bibitem[Journel and Huijbregts(1978)]{journel1978mining}
Andre~G Journel and Charles~J Huijbregts.
\newblock \emph{Mining geostatistics}, volume 600.
\newblock Academic press London, 1978.

\bibitem[Kimura(1980)]{kimura1980statistical}
Akira Kimura.
\newblock Statistical properties of random wave groups.
\newblock In \emph{Coastal Engineering 1980}, pages 2955--2973. 1980.

\bibitem[Lindeberg(1998)]{lindeberg1998feature}
Tony Lindeberg.
\newblock Feature detection with automatic scale selection.
\newblock \emph{International journal of computer vision}, 30\penalty0
  (2):\penalty0 79--116, 1998.

\bibitem[Longuet-Higgins(1957)]{longuet1957statistical}
Michael~Selwyn Longuet-Higgins.
\newblock The statistical analysis of a random, moving surface.
\newblock \emph{Philosophical Transactions of the Royal Society of London.
  Series A, Mathematical and Physical Sciences}, 249\penalty0 (966):\penalty0
  321--387, 1957.

\bibitem[Longuet-Higgins(1984)]{longuet1984statistical}
Michael~Selwyn Longuet-Higgins.
\newblock Statistical properties of wave groups in a random sea state.
\newblock \emph{Philosophical Transactions of the Royal Society of London.
  Series A, Mathematical and Physical Sciences}, 312\penalty0 (1521):\penalty0
  219--250, 1984.

\bibitem[Malara et~al.(2014)Malara, Spanos, and Arena]{malara2014maximum}
Giovanni Malara, Pol~D Spanos, and Felice Arena.
\newblock Maximum roll angle estimation of a ship in confused sea waves via a
  quasi-deterministic approach.
\newblock \emph{Probabilistic Engineering Mechanics}, 35:\penalty0 75--81,
  2014.

\bibitem[McTaggart(2000)]{mctaggart2000ship}
Kevin~A McTaggart.
\newblock Ship capsize risk in a seaway using fitted distributions to roll
  maxima.
\newblock \emph{Journal of Offshore Mechanics and Arctic Engineering},
  122\penalty0 (2):\penalty0 141--146, 2000.

\bibitem[Mirjalili et~al.(2017)Mirjalili, Jain, and Dodd]{Mirjalili_2017}
S.~Mirjalili, S.~S. Jain, and M.~Dodd.
\newblock Interface-capturing methods for two-phase flows: An overview and
  recent developments.
\newblock \emph{Annual Research Briefs}, page 117–135, 2017.

\bibitem[Mohamad and Sapsis(2018)]{mohamad2018sequential}
Mustafa~A Mohamad and Themistoklis~P Sapsis.
\newblock Sequential sampling strategy for extreme event statistics in
  nonlinear dynamical systems.
\newblock \emph{Proceedings of the National Academy of Sciences}, 115\penalty0
  (44):\penalty0 11138--11143, 2018.

\bibitem[Newmark(1959)]{Newmark_1959}
Nathan~M. Newmark.
\newblock A method of computation for structural dynamics.
\newblock \emph{Journal of the Engineering Mechanics Division}, 85:\penalty0
  67--94, 1959.

\bibitem[Onorato et~al.(2009)Onorato, Cavaleri, Fouques, Gramstad, Janssen,
  Monbaliu, Osborne, Packozdi, Serio, Stansberg,
  et~al.]{onorato2009statistical}
Miguel Onorato, L~Cavaleri, S~Fouques, O~Gramstad, Peter~AEM Janssen, Jaak
  Monbaliu, Alfred~Richard Osborne, C~Packozdi, Marina Serio, CT~Stansberg,
  et~al.
\newblock Statistical properties of mechanically generated surface gravity
  waves: a laboratory experiment in a three-dimensional wave basin.
\newblock \emph{Journal of Fluid Mechanics}, 627:\penalty0 235--257, 2009.

\bibitem[Peregrine(1983)]{peregrine1983water}
D~Howell Peregrine.
\newblock Water waves, nonlinear schr{\"o}dinger equations and their solutions.
\newblock \emph{The ANZIAM Journal}, 25\penalty0 (1):\penalty0 16--43, 1983.

\bibitem[Poli et~al.(2007)Poli, Kennedy, and Blackwell]{poli2007particle}
Riccardo Poli, James Kennedy, and Tim Blackwell.
\newblock Particle swarm optimization.
\newblock \emph{Swarm intelligence}, 1\penalty0 (1):\penalty0 33--57, 2007.

\bibitem[Rasmussen and Williams(2006)]{GaussianProcessML}
C.~E. Rasmussen and C.~K.~I. Williams.
\newblock \emph{Gaussian Process for Machine Learning}.
\newblock MIT Press, 2006.

\bibitem[Seyffert et~al.(2016)Seyffert, Kim, and Troesch]{seyffert2016rare}
Harleigh~C Seyffert, Dae-Hyun Kim, and Armin~W Troesch.
\newblock Rare wave groups.
\newblock \emph{Ocean Engineering}, 122:\penalty0 241--252, 2016.

\bibitem[Shigunov(2016)]{shigunov2016probabilistic}
V.~Shigunov.
\newblock Probabilistic direct stability assessment.
\newblock In \emph{Proc. 15th Int. Ship Stability Workshop, 13–15 June,
  Stockholm, Sweden}, 2016.

\bibitem[Shigunov et~al.(2019)Shigunov, Themelis, and
  Spyrou]{shigunov2019critical}
Vladimir Shigunov, Nikos Themelis, and Kostas~J Spyrou.
\newblock Critical wave groups versus direct monte-carlo simulations for
  typical stability failure modes of a container ship.
\newblock In \emph{Contemporary Ideas on Ship Stability}, pages 407--421.
  Springer, 2019.

\bibitem[Shum and Melville(1984)]{shum1984estimates}
KT~Shum and W~Kendall Melville.
\newblock Estimates of the joint statistics of amplitudes and periods of ocean
  waves using an integral transform technique.
\newblock \emph{Journal of Geophysical Research: Oceans}, 89\penalty0
  (C4):\penalty0 6467--6476, 1984.

\bibitem[S{\"o}ding and Tonguc(1986)]{soding1986computing}
H~S{\"o}ding and E~Tonguc.
\newblock Computing capsizing frequencies in a seaway.
\newblock In \emph{Proc. 3rd Int. Conf. on Stability of Ships and Ocean
  Vehicles, Gdansk, Poland}, 1986.

\bibitem[Themelis and Spyrou(2007)]{themelis2007probabilistic}
Nikos Themelis and Kostas~J Spyrou.
\newblock Probabilistic assessment of ship stability.
\newblock volume 115, pages 181--206, 2007.

\bibitem[Umeda et~al.(2004)Umeda, Hashimoto, Vassalos, Urano, and
  Okou]{umeda2004nonlinear}
Naoya Umeda, Hirotada Hashimoto, Dracos Vassalos, Shinichi Urano, and Kenji
  Okou.
\newblock Nonlinear dynamics on parametric roll resonance with realistic
  numerical modelling.
\newblock \emph{International shipbuilding progress}, 51\penalty0 (2):\penalty0
  205--220, 2004.

\bibitem[Weems et~al.(2019)Weems, Belenky, Campbell, Pipiras, and
  Sapsis]{weem2019senvelope}
Kenneth Weems, Vadim Belenky, Bradley Campbell, Vladas Pipiras, and Themis
  Sapsis.
\newblock Envelope peaks over threshold (epot) application and verification.
\newblock In \emph{Proc. 17th Intl. Ship Stability Workshop, Helsinki,
  Finland}, pages 71--79, 2019.

\bibitem[West et~al.(1987)West, Brueckner, Janda, Milder, and
  Milton]{west1987new}
Bruce~J West, Keith~A Brueckner, Ralph~S Janda, D~Michael Milder, and Robert~L
  Milton.
\newblock A new numerical method for surface hydrodynamics.
\newblock \emph{Journal of Geophysical Research: Oceans}, 92\penalty0
  (C11):\penalty0 11803--11824, 1987.

\bibitem[Xiao et~al.(2013)Xiao, Liu, Wu, and Yue]{xiao2013rogue}
Wenting Xiao, Yuming Liu, Guangyu Wu, and Dick~KP Yue.
\newblock Rogue wave occurrence and dynamics by direct simulations of nonlinear
  wave-field evolution.
\newblock \emph{Journal of Fluid Mechanics}, 720:\penalty0 357--392, 2013.

\end{thebibliography}
 
\begin{appendices}
%\numberwithin{equation}{section}
\section{\normalsize{The scale-selection method for wave group detection}}
% rewrite 
%\renewcommand{\theequation}{A\arabic{equation}}

Given any continuous signal $\rho(x)$, the space-scale representation function $S(x,l)$ of $\rho(x)$ is defined by 
\begin{equation}
    S(x,l)=g(x,l)\ast \rho(x),
\end{equation}
where $g(x,l)=(1/\sqrt{2 \pi}l) \exp^{-x^2/(2 l^2)}$ is a Gaussian kernel with scale $l$ at location $x$, and $*$ denotes convolution. 

The scale-selection method is based on the identification of a local minimum of the normalized second derivative of $S(x,l)$, which is defined as
\begin{equation}
    \bar{S}^{(2)}(x,l)=l^2\frac{\partial^2}{\partial x^2} (g(x,l)\ast \rho(x)),
    \label{S2}
\end{equation}
where the normalization factor $l^2$ is used to account for the variation of $S^{(2)}\equiv \partial^2 S/\partial x^2$ due to the length scale of the convolution. Given $\bar{S}^{(2)}(x,l)$, the $L$ and $x_c$ in $\rho(x)$ are captured using Eq.\eqref{opt1}.  

\citet{cousins2016reduced} demonstrate that the direct detection by \eqref{opt1} results in a number of fake groups with low similarity with the $\rho(x)$. To remove these wave groups, a similarity measure (or similarity score) is defined as
\begin{equation}
    C(L,A,x_c)=1-\frac{||\rho_c(x)-A e^{-\frac{(x-x_{c})^2}{2L^2}}||}{||A e^{-\frac{(x-x_{c})^2}{2L^2}}||},
    \label{Cmeasure}
\end{equation}
where $||F||$ represents the $L^2$ norm of a function $F$. The groups with similarity scores
\begin{equation}
    C>0.75
    \label{crit}
\end{equation}
are kept.

%Here we also want to show the close connection between scale-space representation and continuous wavelet transform.
%The second order normalized derivative of $S$ can be written as：

%\begin{equation}
%     \bar{S}^{(2)}(x;L)= \int \rho(u) \frac{1}{\sqrt{L}} G^{(2)}((x-u)/\sqrt{L})du
%\end{equation}
%where $G(y)=(1-y^2)\exp^{-\frac{L^2}{2}}$ is the mother function of Mexican hat wavelet.
%Thus the scale-space representation is actually a $L^\frac{1}{4}$ scaled Mexican continuous wavelet transform. 

\section{\normalsize{ A brief review of Gaussian process regression}}

We consider the task of inferring a random function $r=T(\bm{\theta})$ from dataset $ \mathcal{D}=\{\bm{\theta}^i,r^i\}_{i=1}^{i=n}$ consisting of $n$ inputs ${\rm{\Theta}}=[\bm{\theta}^1,\dots,\bm{\theta}^{n}]$ and the corresponding output ${\rm{R}}=[r^1,\dots,r^{n}]$. The function $T(\bm{\theta})$ is considered as a realization of a Gaussian process, which is specified by its mean and co-variance. A prior, representing our beliefs over all kinds of functions we expect to observe, is placed on function $T$ as jointly Gaussian with a co-variance function $k$ and a mean function $m$: 
\begin{equation}
    T(\bm{\theta}) \sim \mathcal{GP}(m(\bm{\theta}),k(\bm{\theta},\bm{\theta}')).
\end{equation}

If a zero mean is used (as in most cases), the Gaussian processes can then be fully specified by the co-variance function. The implementation of the co-variance is usually through a kernel function
\begin{equation}
    k(\bm{\theta}_p,\bm{\theta}_q) = \sigma_0 \exp{{\frac{-|\bm{\theta}_p-\bm{\theta}_q|^2}{2\delta^2}}} + \sigma_{noise}^2 \delta_{pq},
\end{equation}
where $\delta$, $\sigma_0 $ and $\sigma_{noise}$ are hyperparameters controlling the length scale, convariance amplitude and variance of the white noise, respectively. The values of these hyperparameters are trained by maximizing the likelihood $p(R|\Theta)$.
%The training of GP (optimization of hyperparameters) is implemented by maximizing the log-likelihood $\log p(Y|X)$.  

Given a dataset $\mathcal{D}$, the posterior distribution for a new point $\bm{\theta}^*$ can be deduced by Bayesian rule:
\begin{equation}
    p(r^*|{\rm{R}},{\rm{\Theta}};\bm{\theta}^*) = \frac{p({\rm{R}},r^*|{\rm{\Theta}},\bm{\theta}^*)}{p({\rm{R}}|{\rm{\Theta}})}\sim \mathcal{N}(\overline{r}^*, {\rm{var}}(r^*) )
\end{equation}
with mean and variance:
%\begin{equation}
   % & y^\star|x^\star,Y,X  \sim N(\overline{y}^*,cov(y^*))
%\end{equation}
\begin{align}
    \overline{r}^* &=K(\bm{\theta}^*,{\rm{\Theta}})K({\rm{\Theta}},{\rm{\Theta}})^{-1}{\rm{R}},\\
     {\rm{var}}(r^*)  &=k(\bm{\theta}^*,\bm{\theta}^*)-K(\bm{\theta}^*,{\rm{\Theta}})K({\rm{\Theta}},{\rm{\Theta}})^{-1} K({\rm{\Theta}},\bm{\theta}^*),
\end{align}
where covariance matrixs are: 
\begin{align}
    K({\rm{\Theta}},{\rm{\Theta}})^{ij}&=k(\bm{\theta}^i,\bm{\theta}^j),\\
    K(\bm{\theta}^*,{\rm{\Theta}})^{i} &= k(\bm{\theta}^*,\bm{\theta}^i).
\end{align}

\end{appendices}

\end{multicols}
\end{document}